\DeclareUnicodeCharacter{2212}{-}
\RequirePackage{luatex85}

\documentclass[twocolumn]{aastex63}

\usepackage{tabularx}
\usepackage{comment}
\usepackage{pifont}
\usepackage{natbib}
\usepackage[autostyle]{csquotes}
\usepackage{CJKutf8}
\usepackage{subfigure}

\newcommand{\xmark}{\ding{55}}%

\received{Soon}
\revised{Soon}
\accepted{Soon}
\submitjournal{ApJ}

\graphicspath{{./}{figures/}}

\begin{document}

\title{Peculiar Spectral Evolution of the Type I Supernova 2019eix: A Possible Double Detonation from a Helium Shell on a Sub-Chandrasekhar-mass White Dwarf }

\shorttitle{SN 2019eix: A Peculiar Type I}
\shortauthors{Padilla et al.2022}

\correspondingauthor{Estefania Padilla Gonzalez}
\email{epadillagonzalez@ucsb.edu}

\author[0000-0003-0209-9246]{E. Padilla Gonzalez}
\affiliation{Department of Physics, University of California, Santa Barbara, CA 93106-9530, USA}
\affiliation{Las Cumbres Observatory, 6740 Cortona Dr, Suite 102, Goleta, CA 93117-5575, USA}

\author[0000-0003-4253-656X]{D.Andrew Howell}
\affiliation{Department of Physics, University of California, Santa Barbara, CA 93106-9530, USA}
\affiliation{Las Cumbres Observatory, 6740 Cortona Dr, Suite 102, Goleta, CA 93117-5575, USA}

\author[0000-0003-0035-6659]{J. Burke}
\affiliation{Department of Physics, University of California, Santa Barbara, CA 93106-9530, USA}
\affiliation{Las Cumbres Observatory, 6740 Cortona Dr, Suite 102, Goleta, CA 93117-5575, USA}

\author[0000-0002-7937-6371]{Yize Dong \begin{CJK*}{UTF8}{gbsn}(董一泽)\end{CJK*}}
\affiliation{Department of Physics and Astronomy, University of California, 1 Shields Avenue, Davis, CA 95616-5270, USA}

\author[0000-0002-1125-9187]{D. Hiramatsu}
\affiliation{Department of Physics, University of California, Santa Barbara, CA 93106-9530, USA}
\affiliation{Las Cumbres Observatory, 6740 Cortona Dr, Suite 102, Goleta, CA 93117-5575, USA}

\author[0000-0001-5807-7893]{C.McCully}
\affiliation{Department of Physics, University of California, Santa Barbara, CA 93106-9530, USA}
\affiliation{Las Cumbres Observatory, 6740 Cortona Dr, Suite 102, Goleta, CA 93117-5575, USA}

\author[0000-0002-7472-1279]{C.Pellegrino}
\affiliation{Department of Physics, University of California, Santa Barbara, CA 93106-9530, USA}
\affiliation{Las Cumbres Observatory, 6740 Cortona Dr, Suite 102, Goleta, CA 93117-5575, USA}

\author[0000-0002-0479-7235]{W. Kerzendorf}
\affiliation{Department of Physics and Astronomy, Michigan State University, East Lansing, MI 48824, USA}
\affiliation{Department of Computational Mathematics, Science, and Engineering, Michigan State University, East Lansing, MI 48824, USA}

\author[0000-0001-7132-0333]{M. Modjaz}
\affiliation{Department of Astronomy, University of Virginia, Charlottesville, VA 22904, USA}

\author[0000-0003-0794-5982]{G. Terreran}
\affiliation{Las Cumbres Observatory, 6740 Cortona Dr, Suite 102, Goleta, CA 93117-5575, USA}

\author[0000-0003-2544-4516]{M. Williamson}
\affiliation{Department of Physics, New York University, New York, NY 10003, USA}

\begin{abstract}
We present photometric and spectroscopic data for the nearby Type I supernova (SN Ia) 2019eix (originally classified as a SN Ic), from its discovery day up to 100 days after maximum brightness. Before maximum light SN 2019eix resembles a typical SN Ic, albeit lacking the usual \ion{O}{1} feature. Its lightcurve is similar to the typical SN Ic with decline rates of ($\Delta M_{15,V}= 0.84$) and absolute magnitude of $M_{V}= -18.35$. However, after maximum light this SN has unusual spectroscopic features, a large degree of line blending, significant line blanketing in the blue ($\lambda < 5000$\AA), and strong Ca II absorption features during and after peak brightness. These unusual spectral features are similar to models of sub-luminous thermonuclear explosions, specifically double-detonation models of SNe Ia. Photometrically SN 2019eix appears to be somewhat brighter with slower decline rates than other double detonation candidates. We modeled the spectra using the radiative transfer code TARDIS using SN 1994I (a SN Ic) as a base model to see whether we could reproduce the unusual features of SN 2019eix and found them to be consistent with the exception of the \ion{O}{1} feature. We also compared SN 2019eix with double detonation models and found them to match the observations of SN 2019eix best, but failed to reproduce its full photometric and spectroscopic evolution.
\end{abstract}



\section{Introduction}\label{sec:intro}

As the number of supernova discoveries has grown over time, previously rare or unseen supernovae are being found in increasing numbers. Often their classification can be unclear and their properties may overlap between different classes, making these SNe unique. Consequently, new discoveries have shed light on the diversity of supernovae, and all-sky surveys have played a crucial role in this. For instance, all-sky surveys have discovered interacting SNe with different Circumstellar Material (CSM) compositions, including SNe Ibn \citep{Hosseinzadeh17} and SNe Icn \citep{Gal-Yam2022, Pellegrino2022}. Similarly, even the “well understood” SNe Ia observations have shwown peculiarity in observations that point to  diverse explosion mechanisms. These include “.Ia'' explosions of a helium flash on a white dwarf \citep{Kasliwal10,Poznaski10}, “Sub-Chadrashekar” explosion of He burning followed by second detonation in the core via converging shocks \citep{Fink10, De19}, “Super-Chandrasekhar” explosions of rapidly spinning WD whose limiting mass could exceed the $1.4 \rm M\odot$  \citep{Howell2006, Contreras_2010, Chakradhari14, Parrent16_sc}, and SNe Ia CSM, explosions of SN Ia whose material interacts with the CSM \citep{Raskin&Kasen13, Shen13}.


The progenitor channel that gives rise to SNe Ia is a subject of controversy, with the two primary competing scenarios being the single-degenerate and double-degenerate models. In the single degenerate channel, a White Dwarf (WD) accretes matter from a main sequence or neutron star \citep{Whelan&Ibel1973}. As the WD accretes matter from its companion star, it grows in mass until it approaches a critical limit known as the Chandrasekhar mass. At this point, a runaway nuclear fusion reaction is triggered in the carbon and oxygen-rich material, leading to a powerful thermonuclear explosion \citep{Arnett1969, Nomoto1984}. 

In the double degenerate channel two white dwarfs merge and explode \citep{Iben_tutok84}. Whether or not supernovae are triggered by approaching the Chandrasekhar mass to ignite carbon is debated. Additionally, there exists an alternative explosion mechanisms for WDs by which a WD accretes sufficient He mass (from a He-rich companion) and undergoes ignition, leading to a detonation from the He layer. The detonation of a pure He-shell produces fast and faint transients and they have been used to explain these Calcium-rich transients \citep{Perets10}. Various concerns have been studied, including the conditions for the detonation of the He shell and whether the He detonation can trigger a detonation in the underlying Carbon and Oxygen (CO) core \citep[referred to as the double-detonation scenario, see e.g.][]{Bildsten07,Fink10,Sim12,Shen14,Shen18}.

 Subsequently, various studies have also explored the characteristics of these double detonation events by varying the mass of the He-shell and the CO WD mass \citep{Kromer10, Sim10, Sim12,Polin19}. In double-detonation simulations, the models which approach the limiting case of a bare CO WD detonation are capable of roughly reproducing observations of SNe Ia \citep{Sim10,Shen18}. In contrast, the observables are much different in models which have a thicker He-shell due to the He-rich burning products in the outer ejecta \citep{Sim10,Shen10,Kromer10,Polin19}. As a result, the main spectroscopic observable for these events is a strong suppression in the blue regime. OGLE-2013-SN-079, SN 2016hnk\footnote{although whether it is a double detonation is still debated \citep{Galbany19}}, SN 2016dsg, and 2018byg are all of the double detonation candidates available in literature and from the sample all candidates display such depletion \citep{Inserra15, Jacobson-galan2020, De19, Dong22}. Interestingly, these handful of candidates with the exception of SN 2016hnk have all been found in the outskirts of their host galaxies \citep[with a host type of E or S0;][]{De19,Dong22,Inserra15}, suggesting an older stellar population progenitor for these events.

Subluminous thermonuclear explosions, a subclass of SNe Ia that are 1 to 2 magnitudes dimmer than normal Type Ia's, have been more difficult to understand. \cite{Hachinger09} confirmed that nuclear burning in dim, 91bg-like SNe Ia stops at less advanced stages compared to normal SNe Ia. Despite this finding, the explosion mechanism of subluminous supernovae remains a topic of debate. Some proposed models for subluminous Type Ia supernovae include explosions of sub-Chandrasekhar mass WDs \citep{Fink10} or WDs that undergo a partial deflagration followed by a delayed detonation \citep{Hoeflich02, Mazzali07}.

 Similar to SNe Ia, SNe Ib/c also show a wide range of physical properties. They are core-collapse explosions of massive stars $(> 8M\odot)$ whose outermost layers have been partly or completely stripped of Hydrogen and Helium \citep{Clocchiatti97}. A wide variety of SNe Ib/c exist (for a review on CCSNe, see \citealt{Filippenko} and \citealt{Maryam2019}). Their luminosity and kinetic energy can range more than an order of magnitude, as seen from the overluminous SN 1998bw to the subluminous SN 2004aw \citep{Drout11,Mazzali09}. In addition to the significant luminosity range, the light curves and spectra also are considerably diverse \citep{Liu16, Fremling2018, Williamson19,Barbarino21}. SN 1994I is an example of a normal Type Ic spectroscopically, but stands out due to its fast decline in the R-filter magnitude from maximum time to 15 days later \citep[$\Delta m_{\rm 15,R}=1.5$ mag,][]{Drout11,bianco14,prentice16}. Furthermore, some SNe Ic are characterized by broad features (Type Ic-BL). This type is sometimes associated with gamma-ray bursts (GRBs); see \cite{Woosley&Bloom06} for a review. SNe Ic-BL are spectroscopically distinct in terms of their extreme velocity values that cause the spectral features to blend \citep{Modjaz}.



In this paper, we discuss the peculiar object SN 2019eix. This object is unique because it has many similarities to two distinct classes (e.g., classes: Type Ic and peculiar Type Ia). We will model this object to attempt to find its progenitor and explosion mechanism, more specifically whether it is a core-collapse or a thermonuclear event. The paper is arranged as follows. In Section \ref{sec:discovery}, we discuss the discovery and observations. In Section \ref{sec:Analysis}, we discuss reddening, light curve and spectral analysis in detail. In Section \ref{sec:Modeling}, we model the spectra in two different ways: to replicate the core-collapse SN Ic scenario we directly model the spectra using the radiative transfer code TARDIS \citep{Kerzendorf14,kerzendorf_wolfgang_2022_7331855} and to replicate the double-detonation SN Ia scenario we compare the spectra and light curve models from \citet{Kromer10,Sim12, Polin19} to SN 2019eix. Additionally, we use Arnett's model to infer masses of ejecta and $^{56}$Ni, to better understand the progenitor of the supernova. Finally, in Section \ref{sec:discussion} we discuss the possible physical origins and classification of SN 2019eix and conclude in Section \ref{sec:conclusion}.


\section{Discovery and Observations}\label{sec:discovery}
\subsection{Discovery} \label{sec:2.1}
\begin{figure}
\includegraphics[width=0.45\textwidth]{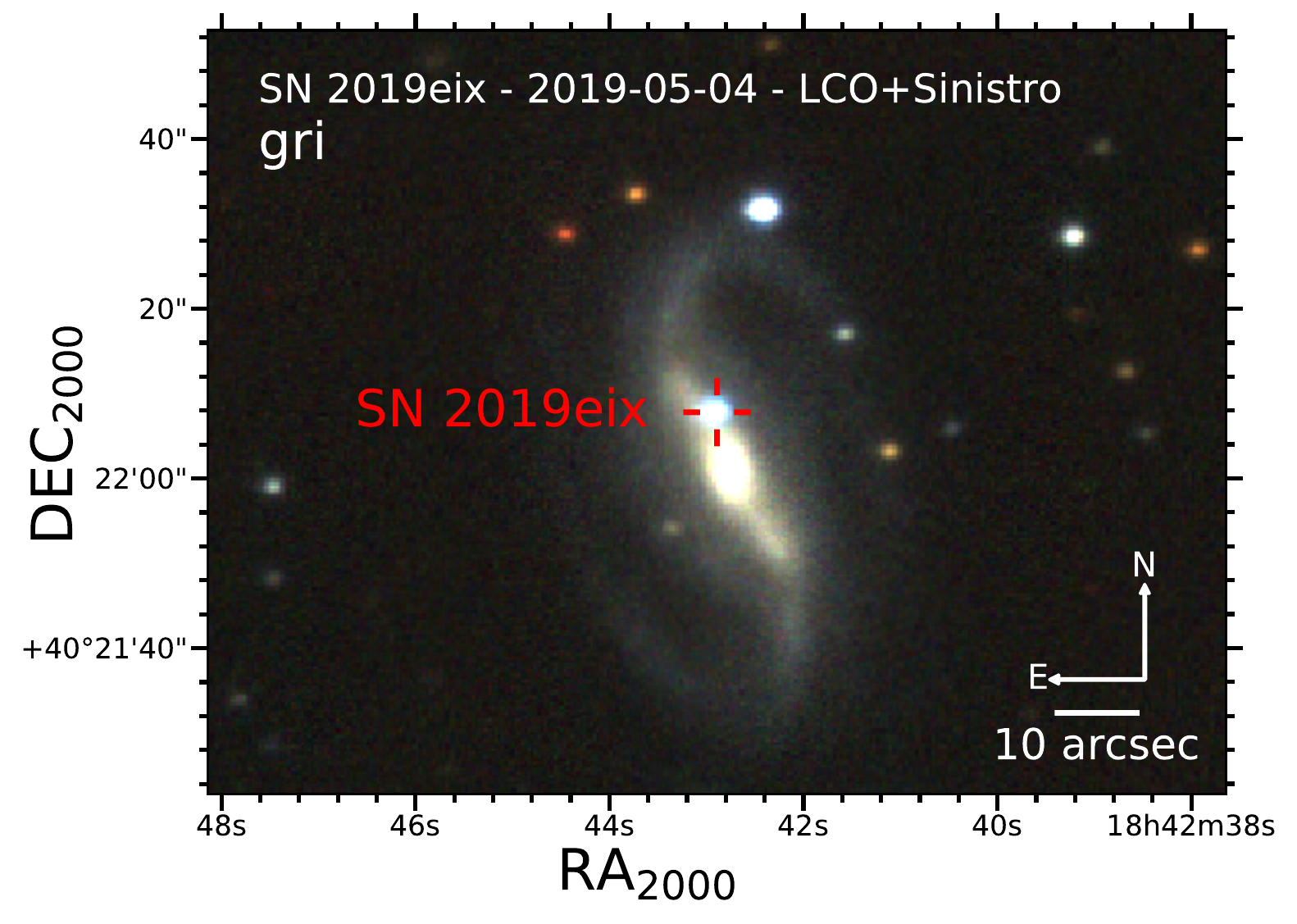}
\caption{gri band staked image of the host of SN 2019eix. The location of the SN is marked in red crosshair.
\label{fig:19eix}}
\end{figure}

 SN 2019eix was discovered by ATLAS on 2019, May 1.59 UT, using the instrument ACAM1 at Haleakalā with a discovery magnitude of 17.0 in the cyan filter \citep{2019TNSTR.687....1T}. The last non-detection of the same object was on 2019, April 21.63 UT with a magnitude of 18.39 (orange filter) from the same telescope. A classification report was provided about a week after its discovery (2019, May 7th), with a classification spectrum taken on 2019, May 5th with the 3 meter Lick Shane reflector telescope suggesting that SN 2019eix was a Type Ic supernova. Due to the lack of early photometry the explosion epoch 2458600 JD was adopted throughout the paper assuming a rise time of 16 days to its maximum light in the r band. A typical Ic rise time falls between 13-16 days \citep{Taddia17} and SN 2019eix appears to be on the wide side for SNe Ic as shown in Figure \ref{fig:lc_comp} thus the choice for 16 days (and the fact that SN 2019eix has similar lightcurves to SNe Ic). The classification spectrum was downloaded from the Transient Name Server (TNS)\footnote{\url{https://www.wis-tns.org/}} and was included in our analysis. The supernova is located at right ascension $18^{h}42^{m}42^{s}.89$ and declination $+40^{\circ}22'07".8$, and is situated in the host galaxy NGC 6695 with a redshift of $z = 0.018303$ \citep{2015ApJ...805...23C}. In Figure \ref{fig:19eix}, we show an image of the field of view, location, and host galaxy of SN 2019eix in the \textit{gri} bands taken with the 1-m telecope of Las Cumbres Observatory \citep[LCO;][]{Brown13} on 2019, May 4th. The distance of the galaxy NGC 6695 is $\approx$83.80 Mpc according to the NASA Extragalactic Database (NED)\footnote{\url{https://ned.ipac.caltech.edu/}} where the Tully-Fisher method was used to calculate the luminosity distance \citep{2007A&A...465...71T}. The Milky Way extinction value $E(B-V) = 0.0604$ was adopted from the \cite{schlafly11} calibration of the \cite{schlegel98} dust maps. 
 

\subsection{Photometry} \label{sec:2.2}
\begin{figure}
\includegraphics[width=0.45\textwidth]{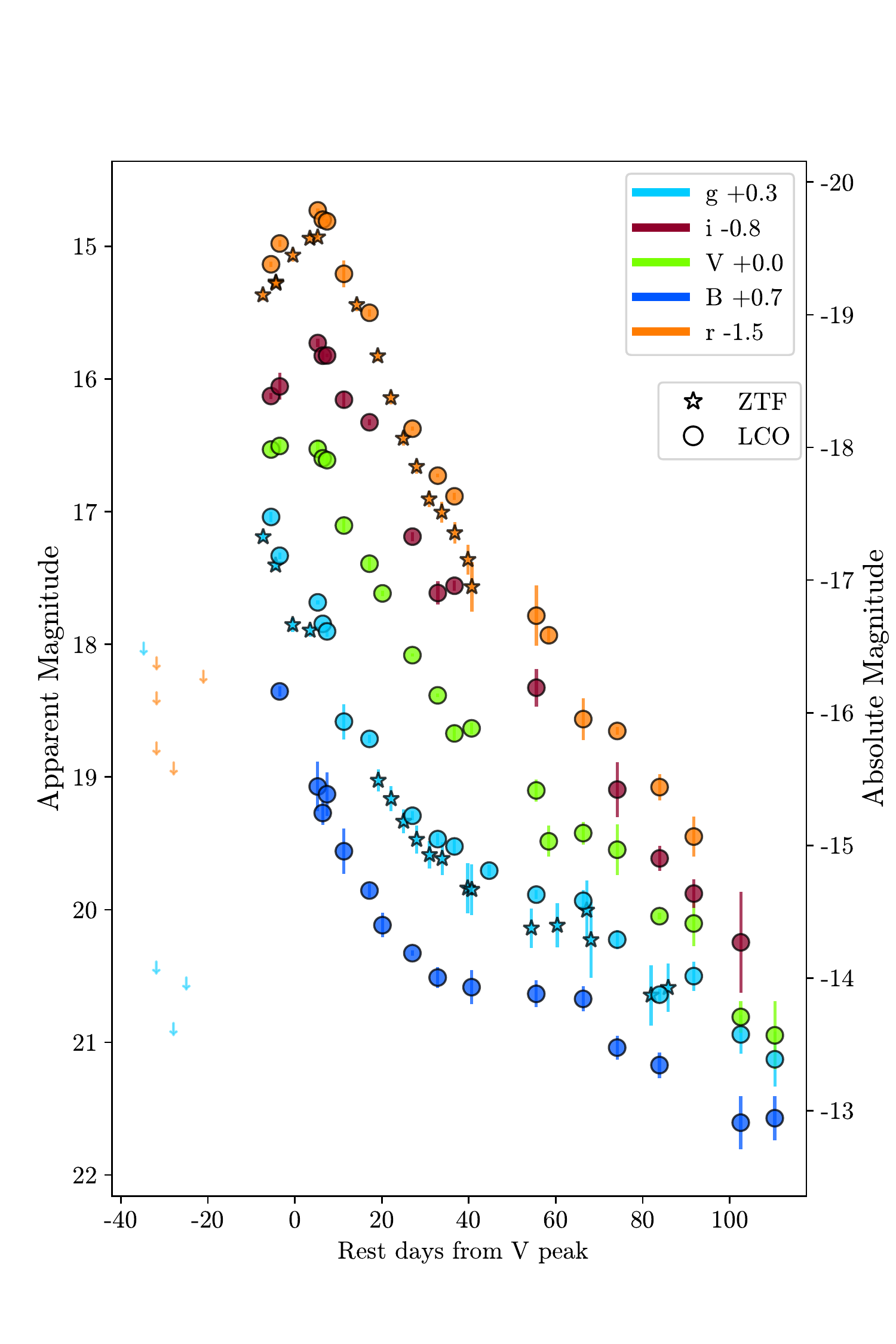}
\caption{Milky Way extinction corrected photometry using $E(B-V)_{\rm{MW}}=0.0604$ as discussed in Section \ref{sec:discovery}. These data are not corrected for host extinction (see Section \ref{sec:3.1} for details on host extinction). The data was obtained from LCO and ZTF, where the data points marked with downward arrows represent the upper limits obtained from ZTF.
\label{fig:19eix_lc}}
\end{figure}

After the discovery, the Global Supernova Project obtained photometry in \textit{BVgri} bands using the 1-m telescopes of LCO . We reduced the data using \texttt{lcogtsnpipe} \citep{Valenti16}. PSF-fitting photometry was performed and the zero points were calculated from the Landolt standard catalogue \citep{Landolt} for the \textit{BV} filters. Additonally, the \textit{gri} filter zeropoints were calculated by incorporating the Sloan magnitudes of field stars \citep{Albareti}. Besides LCO data, the public Zwicky Transient Facility (ZTF)\footnote{\url{https://alerce.online}} data including their upper limits in the g and r bands, were used for comparison and  are plotted in Figure~\ref{fig:19eix_lc}.  

Additional photometric information is presented in Table \ref{table:1}, where it  shows the time of maximum, the peak absolute magnitude, and the decline rates for SN 2019eix. We measured the peak magnitude by fitting a polynomial through the data points and used their maximum value. For the B filter where rising data is not available, we used the peak magnitude from the B band data as an estimate of the maximum light. A comparison of $\Delta m_{15}$ with typical supernovae against SN 2019eix is shown in Table \ref{table:2}, where we can see that SN 2019eix's decline rates fall within the average values for SNe Ic but are somewhat faster in the r and i band \citep[but not unprecedented, see][]{Taddia17}. 


\begin{deluxetable}{ccccc}
\tablewidth{0pt}
\tablehead{
\colhead{Filter} & \colhead{$t_{\rm{max}} (\rm{JD})$} & \colhead{$M_{\rm{max}}$} & \colhead{$\Delta m_{15}$}  
}
\startdata 
B & 2458609.8 & -17.2 & 1.2 \\                
g & 2458607.8 & -18.12 & 1.0 \\
V & 2458612.9 & -18.35 & 0.84 \\
r & 2458615.5 & -18.63 &  0.95 \\
i & 2458616.35 & -18.33 & 0.72 \\
\enddata
\caption{\label{tab:table-name} Time of maximum light, absolute magnitude, and decline rate ($\Delta m_{15}$, the drop in magnitudes between peak and 15 days after peak) for SN 2019eix in each filter of its lightcurve. }
\label{table:1}
\end{deluxetable}

\begin{deluxetable}{ccccc}
\tablewidth{0pt}
\tablehead{
\colhead{Filter} & \colhead{19eix} & \colhead{IIb} & \colhead{Ib} & \colhead{Ic}
}
\startdata 
B & -- & 1.37(0.18) &1.52(0.19) & 1.23(0.59)  \\                
g & 1.0 & 1.10(0.10) & 1.25(0.19) & 1.11(0.45)\\
V & 0.84 & 0.93(0.08) & 1.03(0.19) & 0.90(0.22)\\
r & 0.95 & 0.67(0.08) & 0.75(0.21) & 0.62(0.24)\\
i & 0.72 & 0.53(0.05) & 0.57(0.16) & 0.53(0.17)\\
\enddata
\caption{\label{tab:table-name} $\Delta m_{15}$ per filter for SN 2019eix compared to average values for different SN types, the values in parenthesis correspond to the error. Values were obtained from \citet{Taddia17}.}
\label{table:2}
\end{deluxetable}

\subsection{Spectroscopy} \label{sec:2.3}

\begin{figure}
\plotone{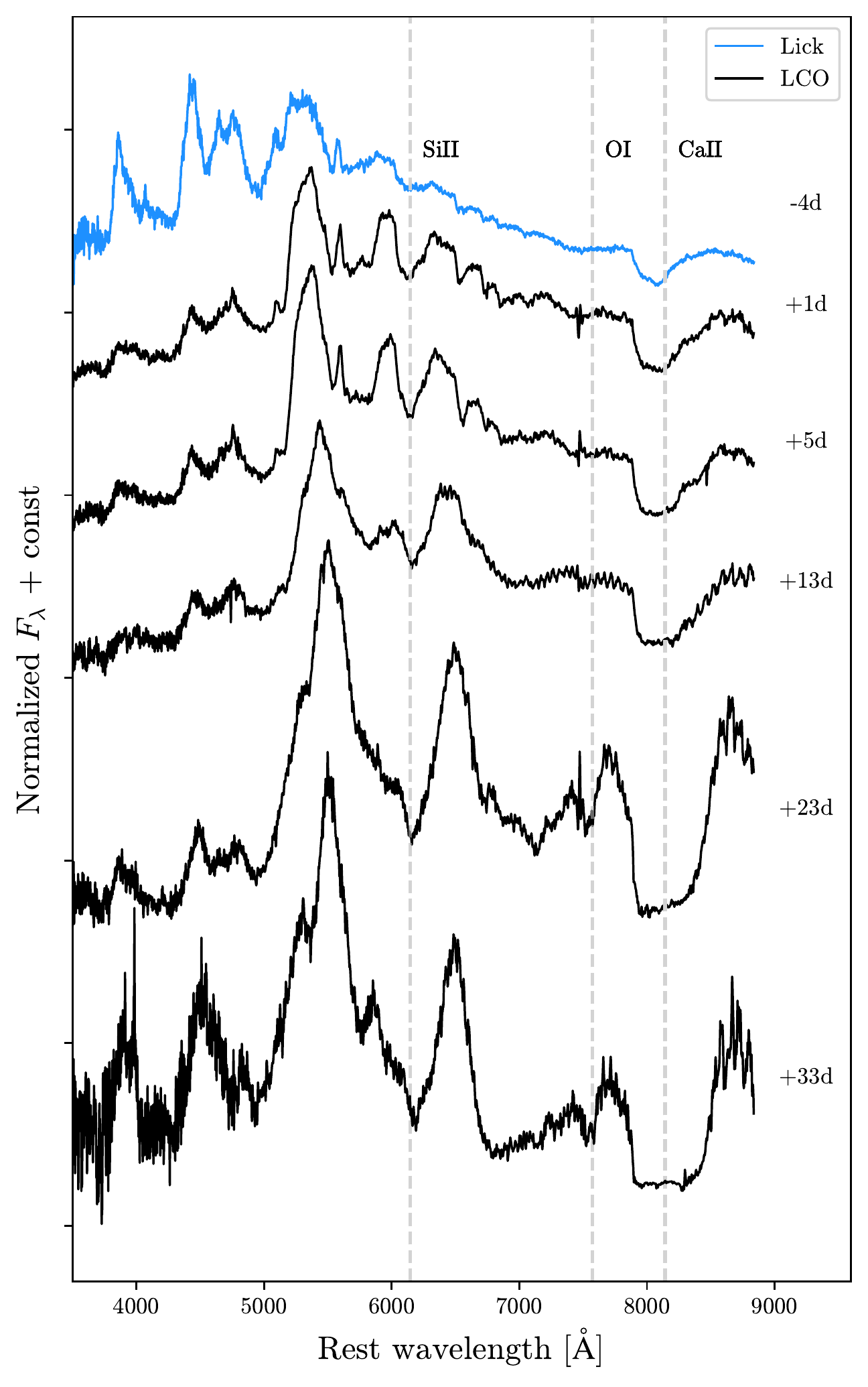}
\caption{ Spectral time series of SN 2019eix. Epochs with respect to V band maximum are included on the right. The dashed lines correspond to the absorption features listed on the top of the figure. The spectra have been corrected for both Milky Way and host extinction (see Section \ref{sec:3.1} for details of host extinction). The last spectrum in the series has an unusual flat-bottomed Ca feature, discussed more in Section \ref{sec:3.2}.
\label{fig:19eix_spec}}
\end{figure}

As part of the Global Supernova Project follow-up, LCO obtained a spectroscopic series using the FLOYDS spectrograph mounted on the 2m Faulkes Telescope North in Haleakalā, Hawai'i. The spectra were reduced as described in \cite{Valenti14}. We also use the classification spectrum posted in TNS (described above); the KAST 3m Lick Shane Reflector took the spectrograph on 2019, May 5.0 UT. All spectra were corrected for Milky Way reddening ($E(B-V) = 0.0604$) and host corrected by measuring the equivalent width of the NaID doublet as described in section \ref{sec:3.1} and shown in Figure ~\ref{fig:19eix_spec}.

\section{Analysis}\label{sec:Analysis}
\subsection{Reddening and Light Curves} \label{sec:3.1}
We investigated two techniques for determining host galaxy reddening which produced different results as shown in Table \ref{table:3}. 
 In the first method, the equivalent width of the sodium line Na ID was measured to be $EW_{\rm Na I D}= 0.139[\rm \AA]$ from IRAF's\footnote{\url{https://iraf.net/}} task \texttt{splot}, using the equation $A_{\rm V}^{\rm host}$ = 0.78($\pm$ 0.15)$\times$ $EW_{\rm NaID}[\rm \AA]$ from \cite{stringer18}. This gives  $A_{\rm V}^{\rm host} = 0.04 \pm 0.15$ and an excess value of E(B-V) of  0.013 $\pm$ 0.002 assuming a reddening law of $R_{\rm v}$ = 3.1.
 
\begin{figure}
\includegraphics[width=0.45\textwidth]{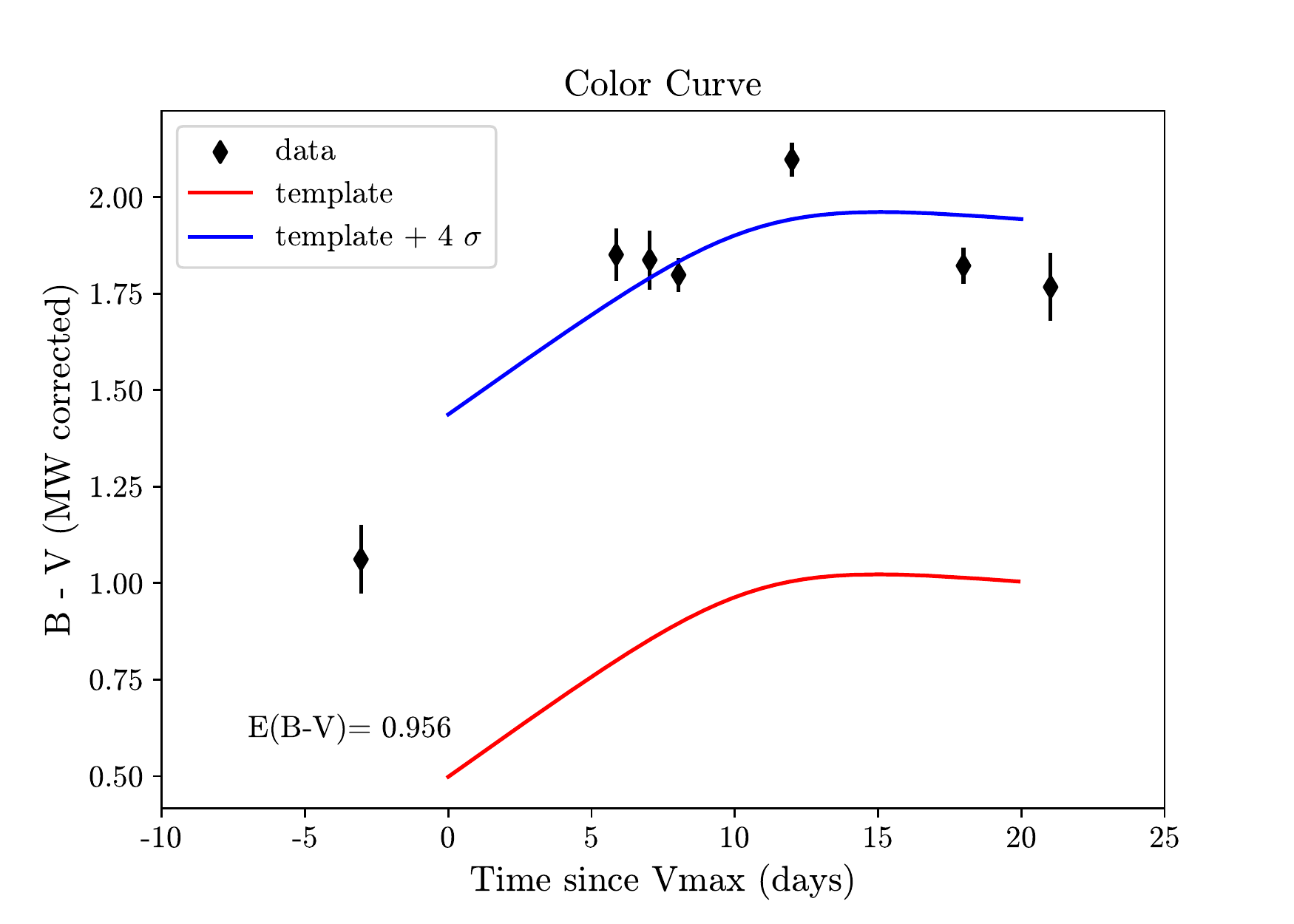}
\caption{Comparison between the color evolution of B-V of SN 2019eix and the Type Ic template. This comparison shows that the template is not a good fit for the color evolution of SN 2019eix. The blue line represents a less likely template which describes the data better, but is still a poor fit.
\label{fig:col_temp}}
\end{figure}

\begin{figure*}
\begin{minipage}{\textwidth}
\centering
\includegraphics[width=0.9\textwidth]{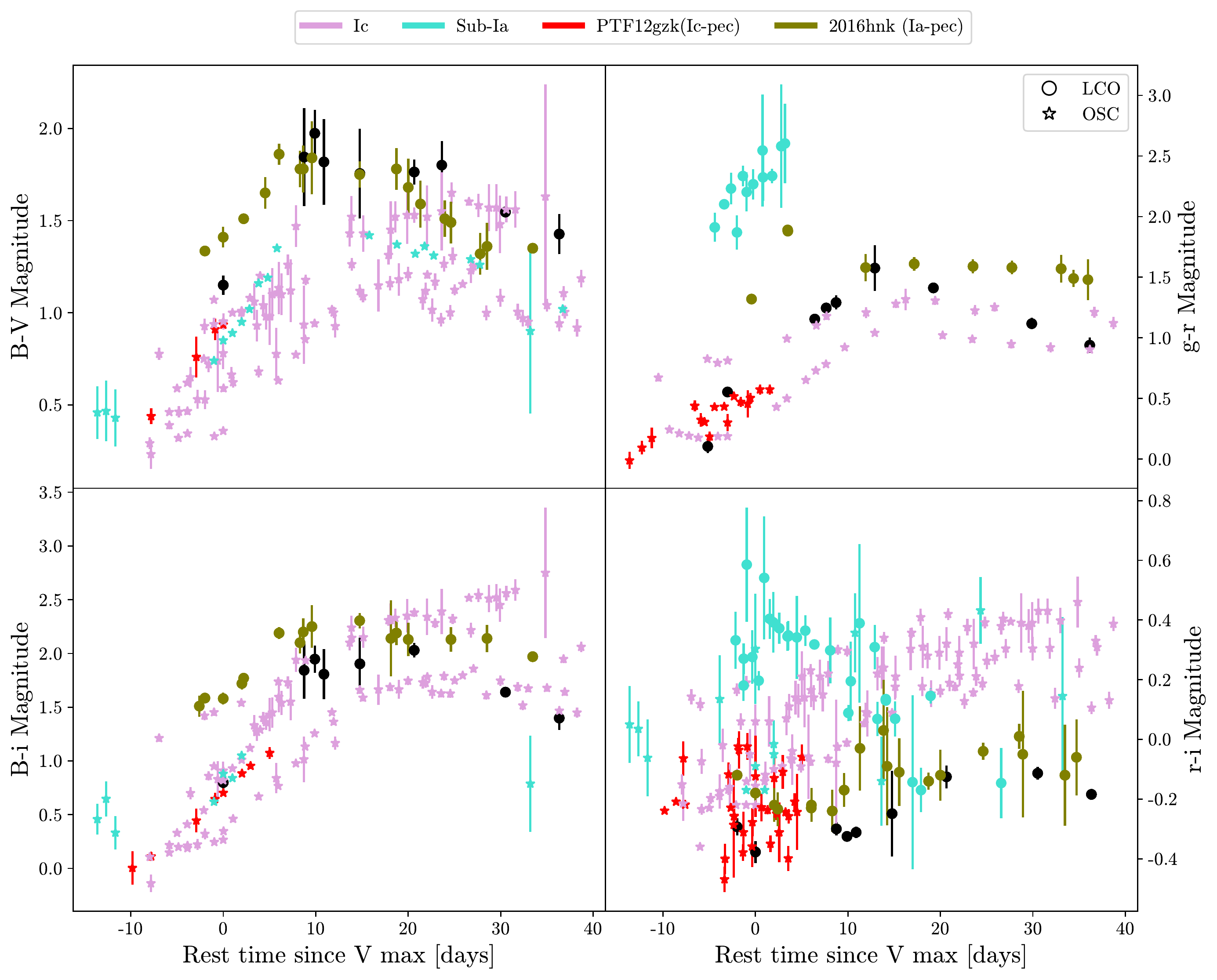}
\caption{The color evolution for SN 2019eix in black along with many SNe Ic (1994I, 2004aw, 2007gr,2004fe, 2004gt) in pink, and sub-luminous Type Ia SNe (91bg, 2005bl, 2016hnk, and 2016dsg) in turquoise were plotted for comparison. Note the difference between SN 2019eix and the SNe Ic. SN 2019eix is much redder than the typical Ic and seems to be more comparable to SN 2016hnk (a double-dentonation candidate).
\label{fig:color_evol}}
\end{minipage}
\end{figure*}

 The second method dealt with comparing the color curves with a template of a typical Type Ic SN \citep{Taddia17}. We considered using this method due to the similarities with Type Ic. Host-galaxy reddening values are estimated through the comparison of observed optical and NIR colors to intrinsic color-curve templates constructed from sub-samples of minimally-reddened CSP-I Stripped Envelope SNe \citep{stringer18}. The templates can be found on the Carnegie Supernova Project (CSP) page \footnote{\url{https://csp.obs.carnegiescience.edu/data/}}. Figure ~\ref{fig:col_temp} shows the color curve B-V of SN 2019eix and the SN Type Ic template against time. The template is not an appropriate template for the data as it does not match the shape of curve due to its peculiarity, it is easier shown when the template is raised by four standard deviations to \textquote{match} the SN 2019eix color data. In Figure \ref{fig:color_evol}, we see that the color evolution of SN 2019eix is inconsistent with SN Ic, meanwhile it exhibits a notable agreement with SN 2016hnk (a double detonation candidate).

The color excess from the weighted difference between the two, is E(B-V) = 0.96. This value is inconsistent with the one measured from the equivalent width of the NaID absorption feature E(B-V) = 0.013. In addition, studies have shown that the V-R for a SNe Ic is about $(V−R)_{\rm V10}$ = 0.26 $\pm$ 0.06 mag where the $V_{\rm 10}$ corresponds to 10 days after V band maximum \citep{Drout11} in contrast SN 2019eix has a value of  $(V −R)_{\rm V10}$ = 0.3612. This points to the possibility that SN 2019eix could be intrinsically more red than a normal SN Ic. This supernova is inconsistent with a reddened normal Type Ic, so we conclude that the red color is primarily from the intrinsic red color of this peculiar event. We therefore adopted the extinction value from the sodium line to correct for reddening.

\begin{figure*}
\begin{minipage}{\textwidth}
\centering
\includegraphics[width=0.95\textwidth]{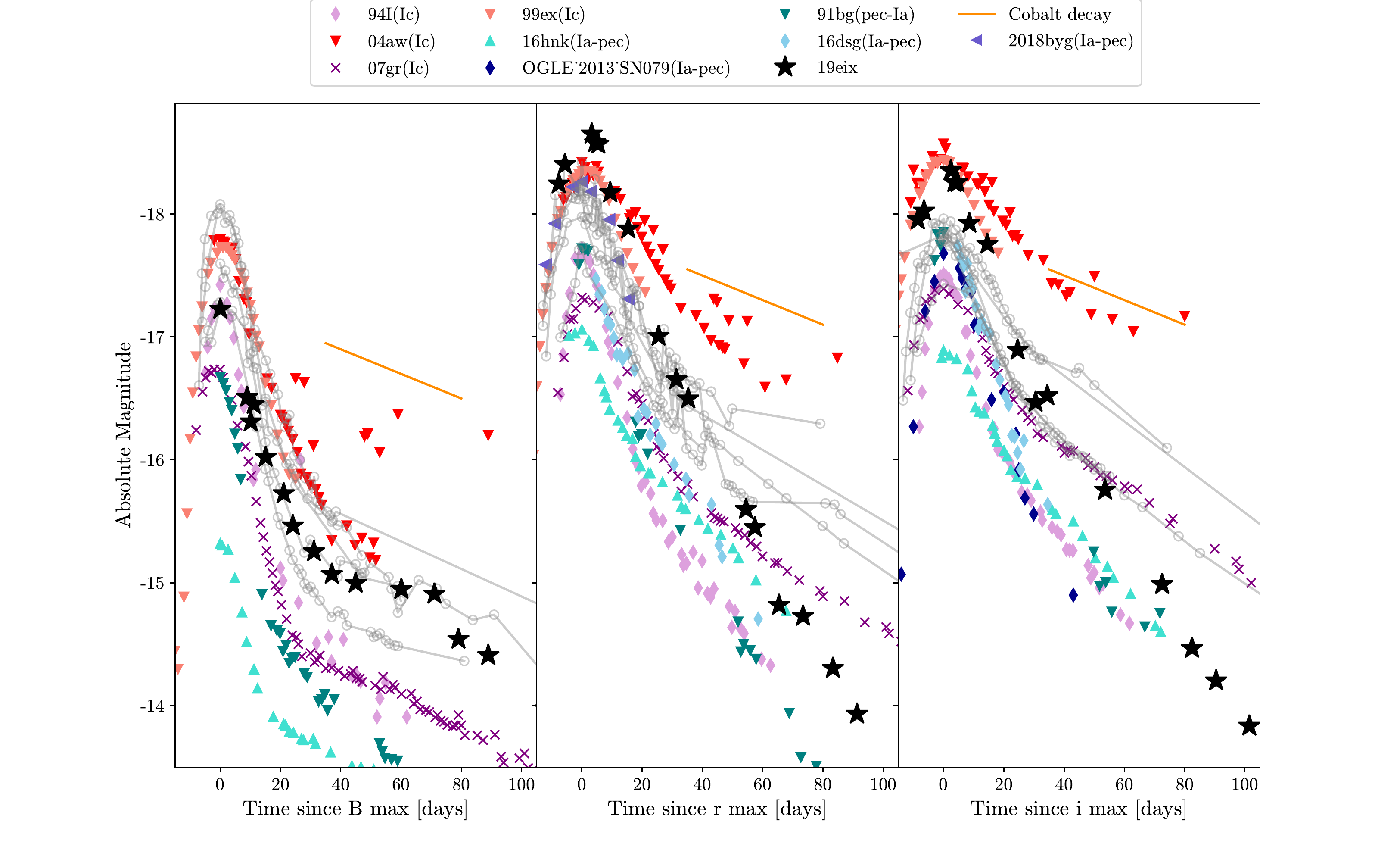}
\caption{Absolute magnitude of SN 2019eix and other Type Ib, Type Ic, and Sub-luminous Type Ia SNe are plotted in the B, r, and i bands. All SNe have been extinction corrected. In this figure the gray (we chose to make some of these SNe in gray for visibility purposes obtained from the Open Supernova Catalogue) and the red/pink lines represent SNe Type Ib/c. We can see how SN 2019eix declines faster than the cobalt decay line and is most similar to Type Ib/c. The double-detonation candidates SN 2016hnk \citep[labeled with \textquote{?} as it is highly debated whether it's a double detonation;][]{Gal-Yam16}, OGLE-2013-SN079 and SN 2016dsg, in the bluer colors appear to be fainter and decline faster than SN 2019eix along with SN 91bg (a sublumnious Type Ia). Generally we notice that SN 2019eix is more similar to Type Ib/c than the double-detonation candidates in terms of light curves.
\label{fig:lc_comp}}
\end{minipage}
\end{figure*}

Figure ~\ref{fig:lc_comp} shows the lightcurves of SNe Ib, Ic, and Ia (with sub-types such as 91bg and double-detonation candidates) along with SN 2019eix in the B , r, and i band. All SNe are Milky Way and host extinction corrected. We observe that the peak absolute magnitude and shape is most similar to Type Ic's including SN 1999ex \citep{Stritzinger_2002} and various others in gray. SN 2016dsg \citep[a thermonuclear He-shell detonation candidate,][]{Dong22} and SN 1991bg \citep[a subluminous SN Ia,][]{Leibundgut93} are the closest in peak magnitude, particularly in the i band (although fainter), with faster evolving light curves than SN 2019eix. For instance, SN 2016dsg has an i band decline rate around 0.077(0.003) mag/day (or $\Delta m_{15}$ = 1.15), as opposed to $\Delta m_{15}$ = 0.72 for SN 2019eix. Generally, from Figure \ref{fig:lc_comp} we see that SNe Ib/c appear to be a better match to SN 2019eix than the thermonuclear SNe in terms of their lightcurve.




\begin{deluxetable}{ccc}
\tablewidth{0pt}
\tablehead{
\colhead{Method} & \colhead{E(B-V)} & \colhead{$\sigma_{E(B-V)}$} 
}
\startdata
Na ID   & 0.013 & 0.002    \\                
Color  & 0.96 & 0.06  \\
\enddata

\caption{\label{tab:table-name}The color excess values for the host using two methods, the equivalent width of the Na ID line listed as \textquote{Na ID} or the weighted subtraction from the color curve of the data and the template listed on the table as \textquote{template}.}
\label{table:3}
\end{deluxetable}

\subsection{Spectral Analysis} \label{sec:3.2}

\begin{figure*}
	\begin{minipage}{\textwidth}
	\centering
    \includegraphics[width=0.9\textwidth]{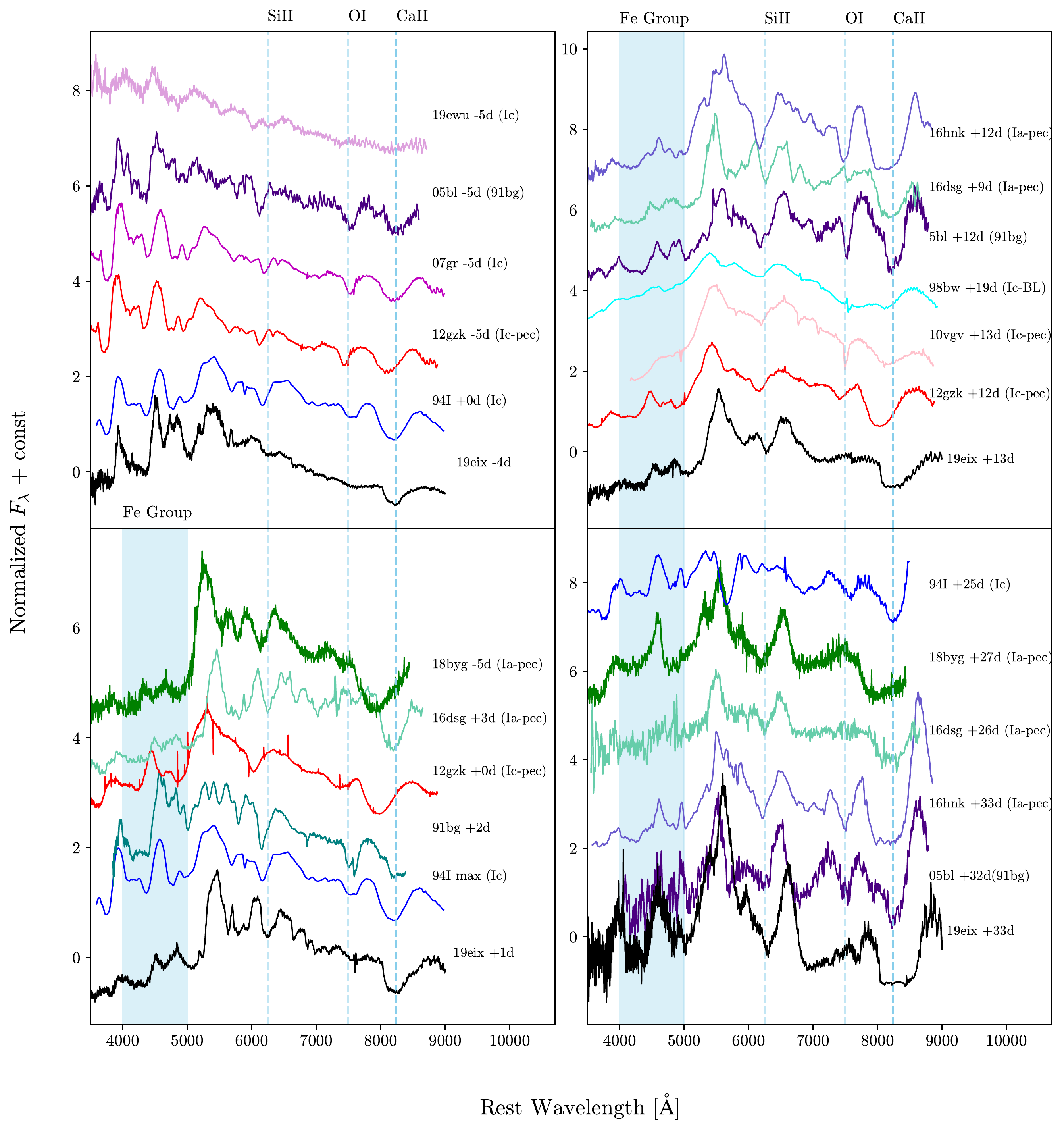}
    \caption{Spectral evolution of SN 2019eix compared with other types of supernovae, including Type Ic and subluminous Ia. The light blue dash lines show the absorption features created by the respective line labeled at the top of the figure. Note the similarities of SN 2019eix at early epochs to Type Ic's and how it transitions to looking more like a sub-luminous Ia (including 91bg-likes, and He-shell double-detonations (DD) candidates at later times. The top left panel shows SN 2019eix at early epochs before maximum along with various Type Ic. The bottom left panel shows SN 2019eix at maximum light and both of the right panels are after maximum light showing a better match to subluminous SNe Ia.
    \label{fig:spec_comp}}
    \end{minipage}
\end{figure*}

The striking features of the spectral sequence of SN 2019eix are concentrated in the suppression of the blue color $\lambda < 5000$\AA , the absence of \ion{O}{1} feature \citep[This is peculiar for SNe Ic which have been shown to generally have a strong \ion{O}{1} 7774 feature, however an exception has been observed;][]{Fremling2018,williamson22}, and the strong evolution of \ion{Ca}{2} feature. In Figure~\ref{fig:spec_comp}, we plot SN 2019eix along with various types of SNe that match spectroscopically at different epochs. We see that the early spectrum at -5 days before maximum matches SNe Ic  \citep[e.g. SN 1994I and SN 2007gr;][]{Gal-Yam16} with the exception of the lack of an \ion{O}{1} absorption feature. We also compared SN 2019eix with SN 2019ewu, a peculiar Type Ic supernova that, like SN 2019eix, lacks an \ion{O}{1} absorption feature. However, SN 2019ewu displays distinct characteristics at this epoch, including an extreme blue continuum and high absorption velocities, which are not typically observed in SNe Ic \citep{williamson22}. These peculiarities are outlined and discussed in greater detail in Section \ref{sec:3.4}. We refrain from comparing SN 2019ewu at later phases to SN 2019eix due to the significant observational differences already mentioned and the reappearance of the \ion{O}{1} feature in the SN 2019ewu spectra.

At -5 days we see strong emission features in the blue between 4000 to 5000 $\rm \AA$ and a less pronounced \ion{Si}{2} feature, much less pronounced than the 91bg-like SN 2005bl. Moreover, as demonstrated in Figure \ref{fig:spec_comp}, the characteristics observed between 5500 and  6500 $\rm \AA$ do not align with those of SNe Ic. Furthermore, as indicated in Figure \ref{fig:tardis}, this particular feature is primarily attributed to Si II. Generally, before maximum SN 2019eix looks like SN 2005bl, but the similarities are not as apparent as they are to SNe Ic.  

At maximum light, the 3500 to 5000 $\rm \AA$ region becomes suppressed making it peculiar compared to the average Ic. SN 2019eix appears to transition in appearance to SN 2018byg a double-detonation candidate and to the fast moving Type Ic PTF12gzk \citep[although the photospheric velocity is much slower than PTF12gzk, having photospheric velocities of $\rm v_{ph}$ $\approx$ 10,500 km/s and $\rm v_{ph}$ $\approx$ 15,300 km/s at peak, respectively;][]{Ben-Ami}. Around 13 days after maximum, the suppression in the blue along with the absence of \ion{O}{1} feature persists. We observe PTF12gzk to evolve like SN 2019eix up until this epoch in terms its general shape (initially resembling a Type Ic and evolving with suppressed features in the blue). Nevertheless, we also noticed key differences including a \ion{O}{1} feature, a smoother profile, and a large blueshift most apparent in the \ion{Ca}{2} and \ion{Si}{2} absorption features in PTF12gzk. Similarly, the double-detonation candidate SN 2016hnk and SN 2005bl (a 91bg-like) also appear to have strong resemblance (however, SN 2005bl does not have as strong of a suppression in the blue), with the exception of the \ion{O}{1} feature. 

The epochs 22 and 32 days after maximum of SN 2019eix seem to resemble 91bg-like and double-detonation Ia's as shown in Figure \ref{fig:spec_comp}, but with even stronger and more blended features. Additionally, the \ion{Ca}{2} feature becomes so broad and large that it even appears to be saturated at 33 days after maximum. We also notice a weak \ion{O}{1} feature previously not observed in the earlier spectra. 

\begin{figure*}
	\begin{minipage}{\textwidth}
	\centering
    \includegraphics[width=0.90\textwidth]{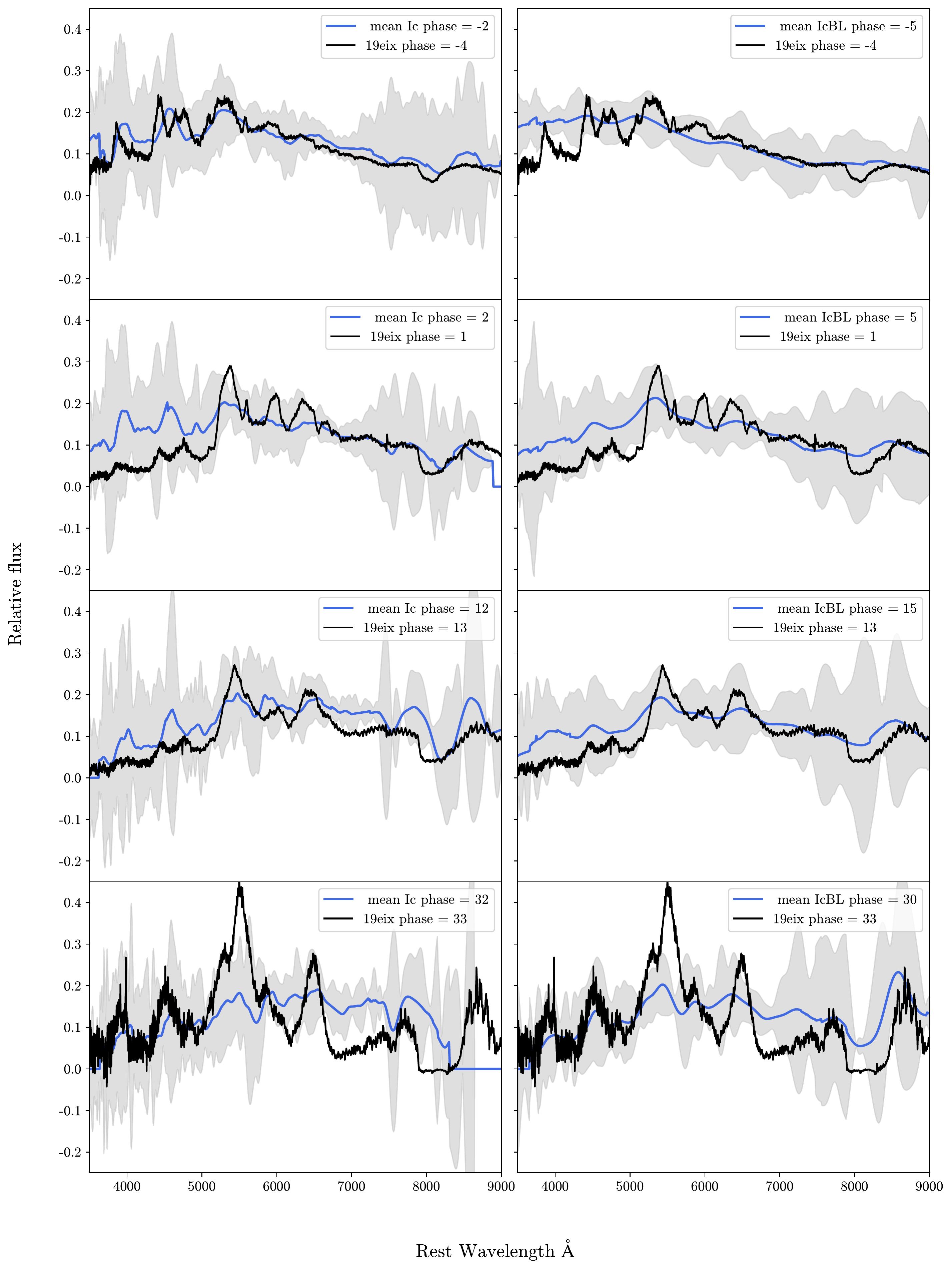}
    \caption{Spectral evolution of SN 2019eix compared with the mean spectra of Type Ic on the left and Ic-BL on the right. The light grey corresponds to one standard deviation from the mean. Although SN 2019eix is not a good match to the templates at later epochs, we do see that before maximum it agrees with the mean Type Ic (on the left panels). After maximum light the shape of SN 2019eix agrees slightly better with Ic-BL than a Ic, but the features for SN 2019eix are sigificantly stronger and more suppressed in the blue. In general, we observed that SN 2019eix is well out of the norm as it is much redder in comparison to both mean spectra for Type Ic an Ic-BL.
    \label{fig:mean_spec_comp}}
    \end{minipage}
\end{figure*}

The spectral evolution presented in Figure ~\ref{fig:spec_comp} does not fully match a typical Type Ic. This is illustrated further when plotting SN 2019eix spectra with the mean spectra of SN Ic illustrated in Figure~\ref{fig:mean_spec_comp}. The templates were used from the NYU SN group github page\footnote{\url{https://github.com/nyusngroup/SESNtemple/tree/master/MeanSpec/}} \citep{Modjaz}. The grey shaded area  represents one standard deviation from the mean at various epochs. These standard deviation regions quantify the spectral diversity within the SN Ic class and facilitate the identification of anomalous features in SN 2019eix. From Figure ~\ref{fig:mean_spec_comp}, the early spectrum is near the mean Type Ic but not at later epochs. SN 2019eix reaches the upper limits of the standard deviation from the mean template and in some cases falls outside the one standard deviation regime. This along with the lack of \ion{O}{1} 7774 feature shows that SN 2019eix is not a typical SN Ic spectroscopically.

\subsection{Pseudo-Bolometric Luminosity} \label{sec:3.3}
To estimate the pseudo-bolometric luminosity we used the technique detailed in \citet{Howell09}. We warped spectral templates for Type Ic SN found from Peter Nugent's page\footnote{\url{https://c3.lbl.gov/nugent/nugent_templates.html}} so that it matched our photometry. We integrated the warped template to get the pseudo bolometric flux at a given epoch where we had data. A bolometric factor was applied in the end to account for the missing flux for those filters with no data and a bolometric luminosity was computed. We also made an effort to determine the bolometric luminosity using a subluminous Type Ia template. Nonetheless, the resulting luminosity showed only a slight alteration, increasing from $3.5\times 10^{42}$ to $3.6\times 10^{42}$. As a result, we decided that the estimate obtained using the Type Ic template was more appropriate and opted to use it in our analysis.

\begin{figure}
\includegraphics[width=0.45\textwidth]{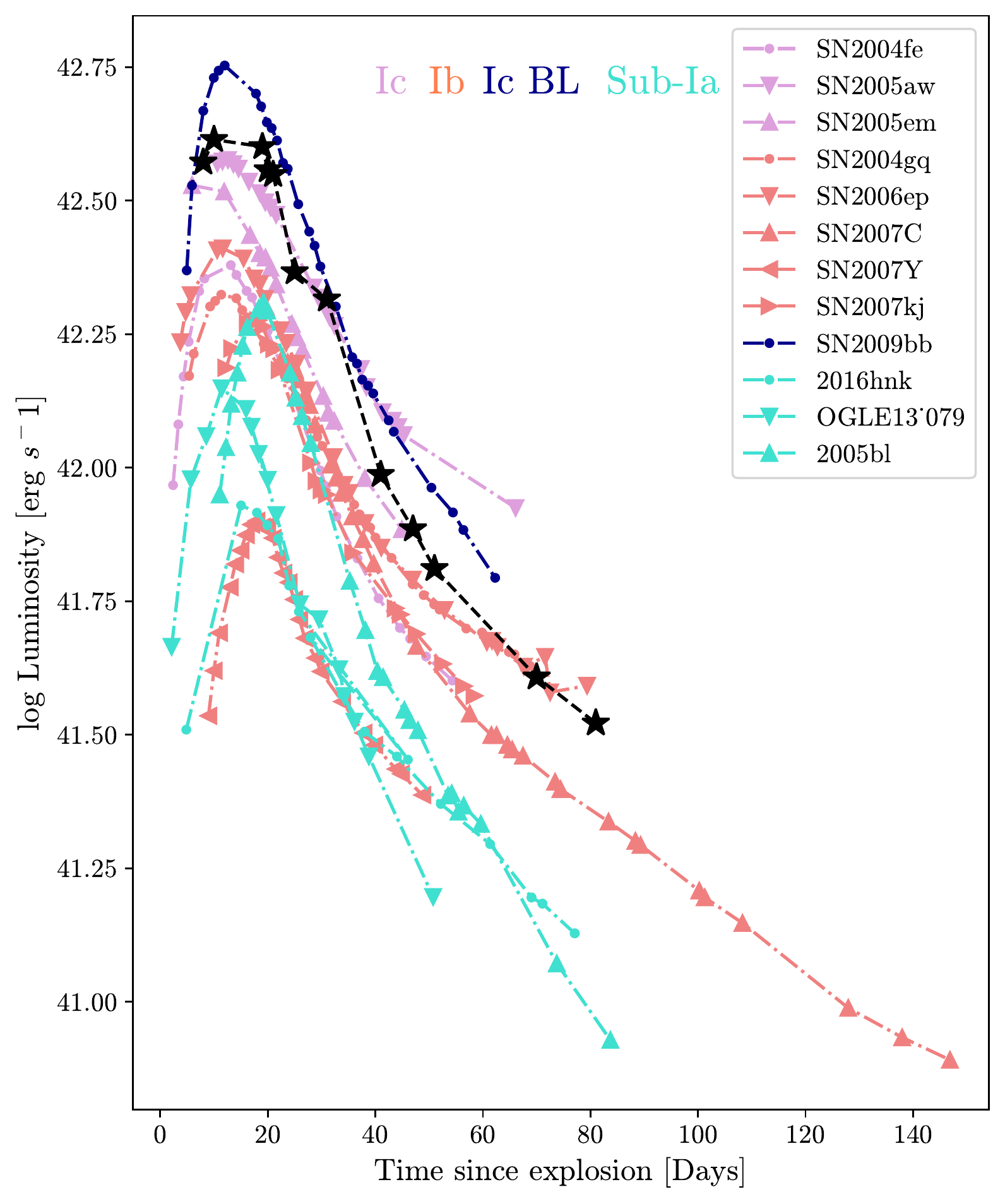}
\caption{Bolometric luminosity of SN 2019eix (in black) plotted with various core-collapse SNe and thermonuclear events. We plotted Type Ic, Ib, Ic-BL, and Sub-luminous Ia in colors pink, coral, dark blue, and turquoise respectively. We note that SN 2019eix is a better match to the bolometric luminosities of SNe Ib/c. \label{fig:bol_lum}}
\end{figure}

The bolometric luminosity of SN 2019eix is plotted along other SNe Types as shown in Figure \ref{fig:bol_lum}. The luminosity for Type Ib/c SNe was acquired from the Carnegie supernova project's website\footnote{\url{https://csp.obs.carnegiescience.edu/data}}. For the subluminous Type Ia comparison, we used the double-detonation candidates including SN 2016hnk from \cite{Jacobson-galan2020} and OGLE13-079 from \cite{Inserra15}; although whether SN 2016hnk it's a double detonation or a 91bg-like it is still debated \cite{Galbany19}. For the other subluminous Type Ia case we used SN 2005bl, a 91bg-like from \cite{Taubenberger08}. From Figure~\ref{fig:bol_lum}, we can see that the maximum bolometric luminosity for SN 2019eix is about $3.5\times 10^{42}$ erg/s/$\rm cm^{2}$/$\AA$. Comparing SN 2019eix with the various types of SNe in this plot, we see that it is a better match to Type Ic. However, SN 2019eix is more luminous than most Type Ic; while much brighter than a subluminous SN Ia.
\vspace{5mm}

\vspace{5mm}
\subsection{Photospheric Velocities} \label{sec:3.4}
\begin{figure}
\includegraphics[width=0.45\textwidth]{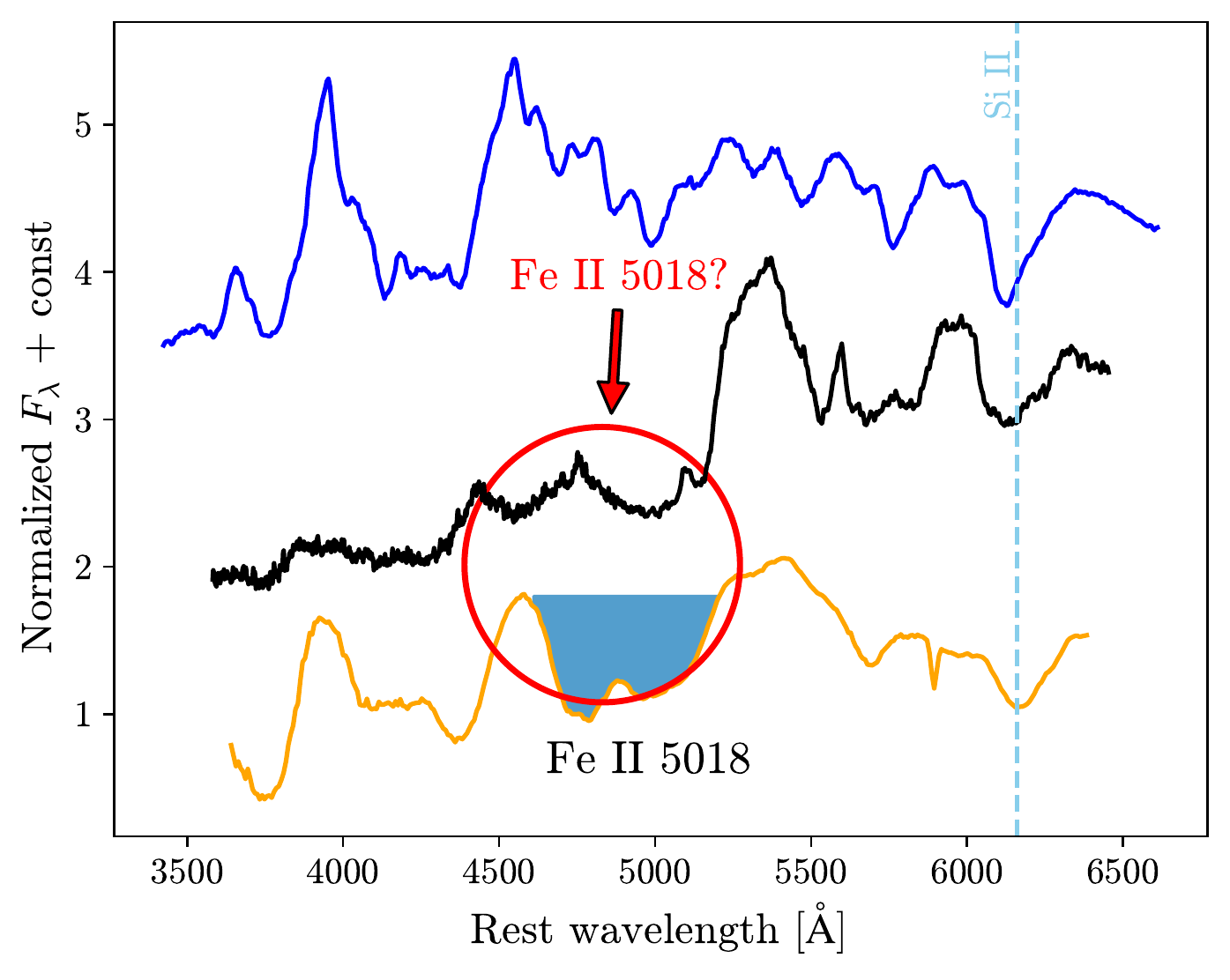}
\caption{Fe II and Si II feature identification, we referred SN 1994I (in orange) from \cite{Hachinger12} to identify the Fe II feature for SN 2019eix (in black). Note that the feature in that region is broad and not obvious to identify. Due to the blending, we used the Si II line to measure the photospheric velocity instead. We also plotted sub-luminous SN 1991bg (in blue) for comparison. 
\label{fig:fe_hidden}}
\end{figure}

\begin{figure}
\includegraphics[width=0.44\textwidth]{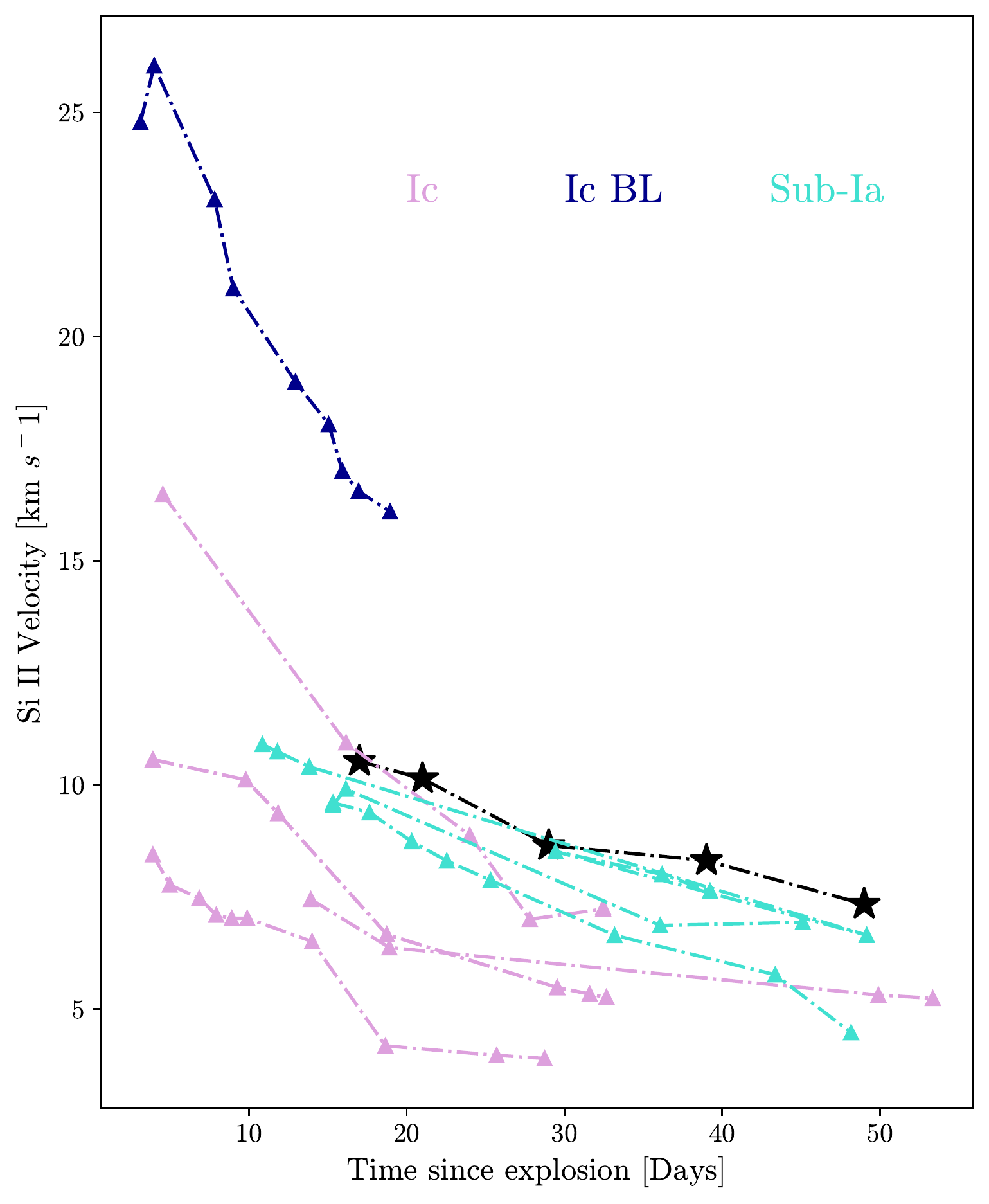}
\caption{ The photospheric velocity is plotted against time for SN 2019eix (in black) and various core-collapse and thermonuclear SNe. We plotted Type Ic, Ic-BL, and sub-luminous Ia in colors pink, dark blue, and turquoise respectively. The velocities were measured using the \ion{Si}{2} feature. Note how SN 2019eix decline rates are most similar to the sub-luminous Ia suggesting similar evolution, but still within range of Type Ic.  
\label{fig:phos_vel}}
\end{figure}

An essential property of the physics of the ejecta used to measure important physical parameters such as the ejecta and $^{56}$Ni mass, is the photospheric velocity. One approach to estimate the photospheric velocity is by measuring the \ion{Fe}{2} 5018 line in the spectra. Because the \ion{Fe}{2} in our spectra was blended and difficult to identify as shown in Figure~\ref{fig:fe_hidden}, we used the \ion{Si}{2} 6355 line instead. Figure~\ref{fig:phos_vel} shows a comparison of the velocity with other types of SNe, including Ic, Ic-BL, and subluminous SNe Ia. SN 2019eix seems to have velocities slightly above a typical SN Ic and be slower evolving. The data from Figure~\ref{fig:phos_vel} was taken from \citet{Ben-Ami}, \citet{Gutierrez2021}, \citet{Taddia16} for the SN Ic and SN Ic-BL. For the sub-luminous SN Ia, we used SN 2016hnk \citep[Calcium-rich transient from a proposed He detonation;][]{Jacobson-galan2020}, SN 2018byg \citep[double-detonation He-shell Type Ia candidate;][]{Jacobson-galan2020}, and SN 2005bl a 91bg-like \citep{Taubenberger08}.  It is important to notice that different absorption features can give different velocities and in some cases the identification of \ion{Si}{2} 6355 line can be problematic \citep{Parrent}. Nonetheless, from Figure ~\ref{fig:phos_vel} we can see that the \ion{Si}{2} 6355 velocities seem to decline at similar rates to the sub-luminous SNe. We also notice that Type Ic SNe decline faster and seem to have lower velocities than SN 2019eix.

\vspace{5mm}

\vspace{5mm}

\section{Modeling} \label{sec:Modeling}

\vspace{5mm}
\subsection{Arnett models} \label{sec:4.1}
Using the pseudo-bolometric luminosity calculated from Section \ref{sec:3.3}, we fit Arnett's model presented by \citet{Arnett82} and \cite{Cano13}, see Cano's equation \ref{eq:1}. We attempt to estimate physical parameters such as the mass of the ejecta and the mass of $^{56}$Ni synthesized in the explosion.

\begin{small}
\begin{eqnarray}\label{eq:1}
L(t) = M_{\text{Ni}}{\rm e}^{-x^2} \times {\left((\epsilon _{\text{Ni}} - \epsilon _{\text{Co}}) \int _{0}^{x}A(z)\,\mathrm{d}z+ \epsilon _{\rm Co}\int _{0}^{x}B(z)\,\mathrm{d}z\right)} \nonumber\\
\end{eqnarray}
\end{small}
where,

\begin{equation}
A(z)=2z\exp ^{-2zy+z^2}, B(z)=2ze^{-2zy+2zs+z^2}
\end{equation}

\begin{equation}
\tau _{m} \approx \left(\frac{\kappa }{\beta c}\right)^{1/2} \left(\frac{{M_{\rm ej}}}{{v_{\rm ph}}}\right)^{1/2}
\end{equation}

and $x\equiv$ $t/\tau_{m}$,  $y\equiv$ $\tau_{m}$/(2$\tau_{Ni}$) and s $\equiv$ ($\tau_{m}$($\tau_{Co}$ − $\tau_{Ni}$)/(2$\tau_{Co}$ $\tau_{Ni}$)).

The assumptions used here are the same as the stripped envelope case from \cite{Cano13}. The constant opacity $\kappa$ = 0.07 $\rm cm^{2}g^{-1}$ and the constant of integration $\beta$ was set to 13.8. The energy released in one second per gram of $^{56}$Ni and $^{56}$Co are $\epsilon_{\rm Ni} = 3.90\times 10^{10}$ erg $\rm s^{−1} g^{−1}$ and $\epsilon_{\rm Co} = 6.78 \times 10^{9}$ erg $\rm s^{−1} g^{−1}$ \citep{Cappellaro97}, respectively with decay times of $\tau_{Ni}$ = 8.77 d and $\tau_{Co}$ = 111.3 d \citep[][and references therein]{taubenger06}.

The fit is done using data prior to thirty days after explosion (the bolometric luminosity used is shown in Figure \ref{fig:bol_lum}) using $\rm v_{ph} \approx 10,500$ km/s measured from the \ion{Si}{2} feature and an assumed time of maximum of 16 days. From the relationship between $\rm v_{ph}$ and the ejecta energy, the kinetic energy was calculated using $\rm E_{K}/M_{ej} =\frac{3}{10} v_{ph}(t_{max})^{2}$ \citep{Wheeler14}. Table \ref{table:summary_sne} shows the resulting parameters of SN 2019eix from the Arnett's fit, where the estimated ejecta mass is $\approx$ 2.5 $\rm M_{\odot}$ and the nickel mass $\rm ^{56}$Ni is $\approx$ 0.17 $\rm M_{\odot}$. The rest of the values are taken from Table 8 \citep{Taddia17} for comparison, and we notice they agree with a Type Ic supernovae. In contrast, some of the Type Ic SNe with similar spectra had values that were more diverse, showing the extreme range of ejecta masses a Type Ic/Ic-BL can have. SN 1994I Ic had ejecta mass of $\rm M_{\rm ej}$ = 0.5 ± 0.2$\rm M_{\odot}$ \citep{Nicholl}, PTF10vgv Ic-Pec had $\rm M_{\rm ej}$ = 1.5 $\rm M_{\odot}$, and lastly PTF12gzk Ic-pec had $\rm M_{\rm ej}$ = 7.5 $\rm M_{\odot}$ \citep{Ben-Ami}. SN 1994I had fast rise times and therefore less massive ejecta. On the other hand PTF12gzk had long rise times, larger ejecta masses and high ejecta velocities of about 30,000 km/s days after explosion \citep{Ben-Ami}.

From the measured parameters in Table \ref{table:summary_sne} we observe that these masses ($^{56}$Ni and ejecta) and energies of SN 2019eix are more consistent with those of a typical Type Ic, since any thermonuclear event would require a lower mass than the 1.4 M$_{\odot}$ limit. Another method we used to estimate the mass of $^{56}$Ni, is from the tail of the light curve. We used equations 1 and 2 from \citet{Terran16} and fit the equations to our late photometry (past 50 days after maximum) and estimated a $\rm ^{56}$Ni mass of $\approx$ 0.09 $\rm M_{\odot}$. This mass is lower than the one approximated from peak. For our purposes, we will use this value and add a 0.08 $\rm M_{\odot}$ error estimate to include the measured value from peak for consistency. Additionally, to test for inconsistencies we also applied the constant opacity $\kappa$ = 0.1 $\rm cm^{2}g^{-1}$ commonly used for SNe Ia \citep{Lyutykh21}. The resulting mass of ejecta slightly decreased from 2.5 $\rm M_{\odot}$ to 2.1 $\rm M_{\odot}$, bringing it closer to the Type Ia scenario, but it still remains significantly higher than the 1.4 $\rm M_{\odot}$ threshold.

It is important to note a major limitation in applying semi-analytic modeling techniques (Arnett's model) to stripped envelope light curves is the assumption of a constant opacity, $\kappa$, ambiguity regarding the velocity, and lack of early photometry. It was shown by \cite{Dessert16}, how different values of $\kappa$ can lead to different results for the progenitor parameters, and that ultimately the assumption of constant opacity is quite poor for stripped envelope SNe. \citet{Parrent} showed that the \ion{Si}{2} velocity was problematic, while the \ion{Fe}{2} 5018 can be hard to identify as shown in Figure~\ref{fig:fe_hidden}. Therefore there is no true velocity to use as both methods come with uncertainties and the estimates of the ejecta mass highly depend on this velocity. Earlier photometry would have allowed for a more accurate rise time and as a result more accurate ejecta mass estimates from the lightcurve, but lacked these early observations. Additionally, we assumed no $^{56}$Ni mixing, if SN 2019eix is a double-detonation a significant amount of $^{56}$Ni would be synthesized in the surface of the He shell. This is equivalent to some degree of $^{56}$Ni mixing that could possibly affect our estimates of the ejecta and $^{56}$Ni masses. Overall, the $^{56}$Ni mass error is dominated by the error on the SN distance, explosion epoch estimate, the fit uncertainty, and the possible $^{56}$Ni mixing synthesized in the He shell (if SN 2019eix is a double-detonation).

\vspace{5mm}

\begin{table*}
    \begin{center}
    \centering
    \makebox[1 \textwidth][c]{       
    \resizebox{0.8\textwidth}{!}{   
    \hspace*{-9em}
    \begin{tabular}{ccccccccc} 
    
    \hline
    \hline
    
    Supernova &$M^R_{\rm max}$ & $\Delta m_{15} (R)$ & $M_{ \rm ej}$[$M_{\odot}$] & $M_{ \rm Ni}$[$M_{\odot}$] & Spiral Gal \\
    \hline
    19eix & -18.32 & 0.95 &  2.5(1.0) & 0.09(0.05) & \checkmark    \\
    Ic & −18.3(0.6) &  0.73(0.27) & 2.1(1.0) & 0.13(0.04) & \checkmark  \\
    Ic-BL &−19.0(1.1) & 0.6(0.14) & 4.1(0.9) & 0.96(1.09)&  \xmark \\
    91bg like & -16.7 to -17.7(B) & 1.8-2.1 (B)  & 0.5  & 0.05-0.1 & \xmark \\
    Ultra stripped (2005ek) & -17.26(0.15) & 2.88(0.05) & 0.3 & 0.03 & \xmark   \\
    Calcium-rich transient & -15.5 to -16.5 & ?  & 0.4-0.7 &$\approx$ 0.01 & \xmark \\
    Ia-DD (2018byg) & -18.2 & 1.23  & $<$ 1.4 & 0.11 & \xmark  \\
    \hline
    
    \end{tabular}
    } 
    } 

    \caption{\label{tab:table-name} This table summarizes our findings and we check the similarities between SN 2019eix in the R band \citep[we converted r to R using the vega - AB magnitude conversion,][]{Blanton2007}  and the different classes, from this table we can see that Type Ic shares more similarities than the other classes. The inputs with an \textquote{(B)} indicates the value in the B band as the R band is not provided and the values in parentheses are the errors.}
    \label{table:summary_sne}
    \end{center}
\end{table*}

\vspace{5mm}

\subsection{TARDIS models} \label{sec:style}

Due to the complex nature of the spectra of SN 2019eix, we experimented with whether it was possible to reproduce some of the unusual features using a SN Ic as a base model (as originally classified). By employing TARDIS, a Monte Carlo radiative transfer code that tracks the number of photons that propagate through the supernova ejecta, we simulated the spectra of SN 2019eix. The version of TARDIS utilized in this study is TARDIS 2022.11.21 \citep{Kerzendorf14}, which is based on previous works \citep{Abbott85, Lucy93, MazzaliLucy93, Lucy99, Lucy02a, Lucy03}. The user must input the density structure, velocity, and abundance of elements. TARDIS then solves the ionization and excitation states of the plasma assuming a homologous expansion. The evolution in the early photospheric phase is such that the photosphere can be approximated to be optically thick and thus the black body approximation is appropriate. Additionally, TARDIS approximates the photosphere by assuming it's position in velocity space. For the simulations presented in this paper, we use the Kurucz atomic dataset version 1.0 (i.e. line list) for calculating the bound-bound transitions. We also used the initial parameters listed in Table 1. from \citet{Marc} and their input files for SN 1994I that can be found in the Github page \footnote{\url{https://github.com/tardis-sn/tardis-setups/tree/master/2020/2020_williamson_94I}}.


Given that SN 2019eix is inconsistent with the spectral evolution of typical Type Ic, we modeled SN 2019eix spectra using a SN 1994I (SN Ic) spectral model at day 16 as our base model. We adjusted the abundance \citet{Hachinger12}, density, velocity, and temperature of radiation according to different epochs and set the damping constant to 0.5. We changed these parameters in order to see whether the spectral features of SN 2019eix could be reproduced, as discussed below. The epochs modeled include 4 days before maximum, 1 day after maximum, and 13 days after maximum light as shown in Figure~\ref{fig:tardis}. All of our models were well converged in the the temperature of radiation space.

In the Appendix \ref{sec:Appendix}, we plot the changes made to the temperature and density profiles and in Figure \ref{fig:dens_prof_Tardis}, we show the changes made to the abundances per epoch. Note in Figure \ref{fig:tardis} that the models fail to reproduce the \ion{O}{1} region as they are unable to attenuate it. Attempts were made to deplete the \ion{O}{1} feature throughout all epochs, by either raising the temperature (in attempt to ionize the \ion{O}{1}) or lowering the abundance in the outer layers. We found that raising the temperature to ionize the \ion{O}{1} was unable to fully eliminate the absorption feature in the model. Additionally, the temperature increase caused the spectra to gain bluer features that did not match our observations. Lowering the abundance on the other hand (various combinations were attempted including depleting the outer shells, reducing \ion{O}{1} in all shells, and  linearly decreasing the \ion{O}{1} in all shells) did not seem to affect the spectra as shown in Figure \ref{fig:tardis}.

The phase 4 days before maximum light was assumed to be 13 days after explosion in our TARDIS model, with an accompanying photospheric velocity of 12200 km/s (chosen based on the measured \ion{Si}{2} velocity and the velocity values that best-matched TARDIS models to observations). We used the reference model with a slight adjustment in the density and temperature which were all increased. Additionally Ti was increased in the earlier shells and Si was decreased. Although not perfect, this combination yielded the best results that matched our observations. One issue with it was the overprediction of the \ion{Si}{2} feature. Attempts were made to enhance the feature including decreasing its abundance, but failed to reproduce it.

At around maximum light (+1 day), we used a time of explosion from SN 2019eix of 19 days and a photospheric velocity of 11700 km/s. We lowered the density profile from our reference model as the ejecta expands and kept the same temperature from the base model. A slight increase of Ti and Fe, gave the best match to our data. One approach we used to determine which elements to alter in our models was comparing the absorption and emission features from Figure \ref{fig:tardis} with the data. From this Figure, we were able to identify which elements could be responsible for creating particular features. Therefore, altering the abundance of these elements could help achieve the desired results and uncover the tomography of the supernova at specific epochs. When contrasted with the results from \citet{Hachinger12} at 22 days after the explosion, it becomes apparent that SN 2019eix has considerably higher concentrations of Ti and V in its inner layers. On the other hand, the spectral model for 1994I \citep{Hachinger12} at 22 days after the explosion shows more significant amounts of Ni and Fe in the inner layers, indicating a substantially faster evolution rate than that of SN 2019eix.

At phase 12 days after maximum light a time of 29 days after explosion was assumed for the model and a photospheric velocity of 10370 km/s. The density was significantly reduced along with the temperature and the Fe was increased by a factor of 5 from our reference model. When comparing this model to SN 1994I \citep{Hachinger12} at day 30 after the explosion we notice it to be significantly more evolved than ours. Their velocity at this point has dropped by a factor of 6 from day 16 and has significantly more Iron-Group Elements (IGE) and Intermediate-Mass Elements (IME) (such as Fe, Ni and Si, S, respectively) in the inner shells than we observe for our SN 2019eix model. SN 1994I is a fast evolving Ic, whereas SN 2019eix appears to not change much in terms of its abundance or photospheric velocity from before to after maximum light.

From our TARDIS models of SN 2019eix in comparison to \cite{Hachinger12}, we note that the velocity does not evolve as fast as SN 1994I \citep{Hachinger12} drops by a factor of 6 in 14 days, and in SN 2019eix models it drops by a factor of 1.2. The temperature has the opposite effect and appears to drop faster for SN 2019eix than for the \cite{Hachinger12}. Additionally, SN 2019eix also seems to be denser than \cite{Hachinger12} throughout all the epochs. SN 2019eix seems to be a slow evolver in terms of velocity and reaction rates. A possible explanation to these peculiar features we observe in SN 2019eix could be a density and temperature effect, as those were the main parameters we varied the most and appeared to successfully reproduce the spectra as shown in Figure \ref{fig:tardis}.\footnote{All of our model input files can be found on: \url{https://github.com/tardis-sn/tardis-setups/2022/2019eix}}


\begin{figure*}
    \centering
    \subfigure[]{\includegraphics[width=0.45\textwidth]{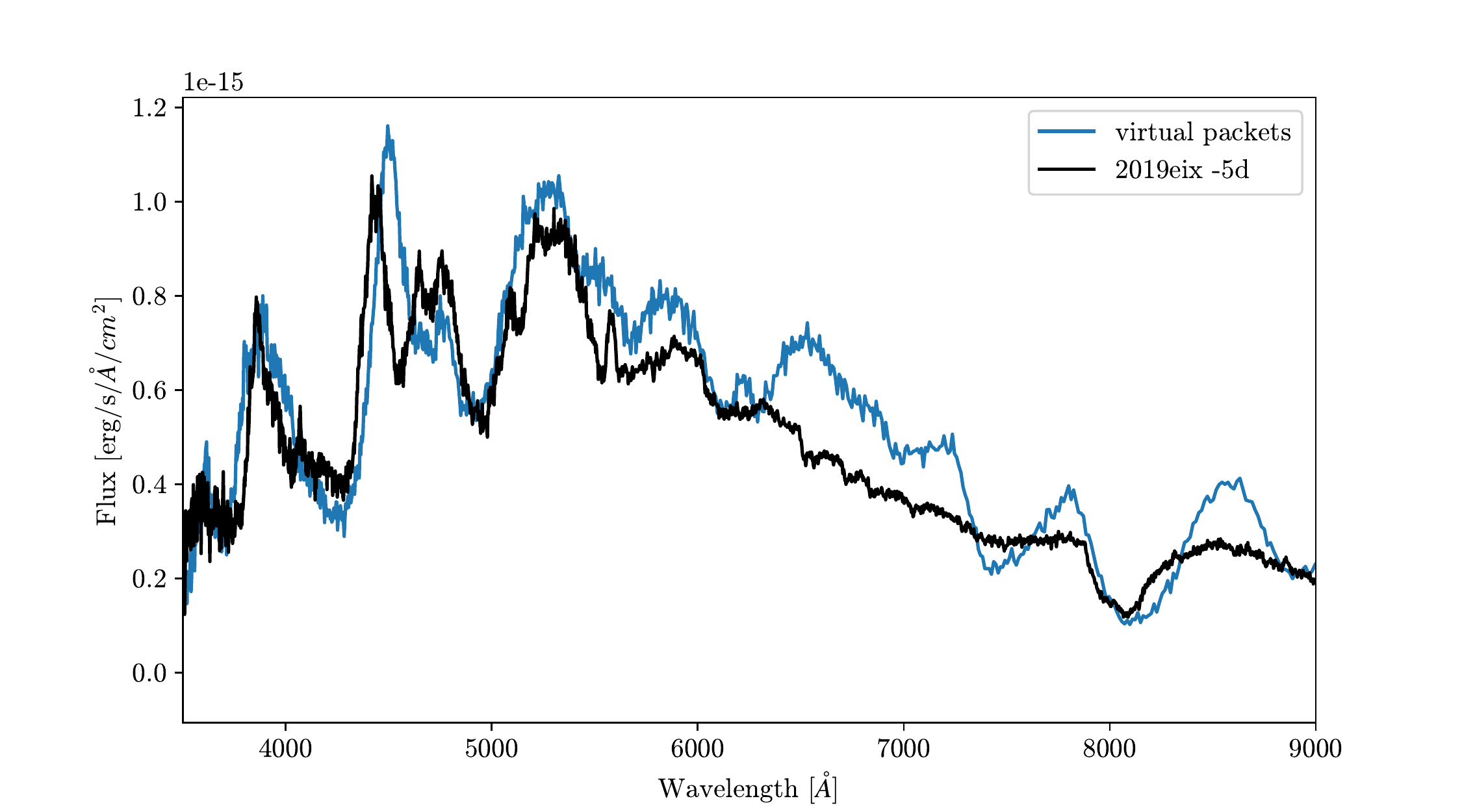}} 
    \subfigure[]{\includegraphics[width=0.53\textwidth]{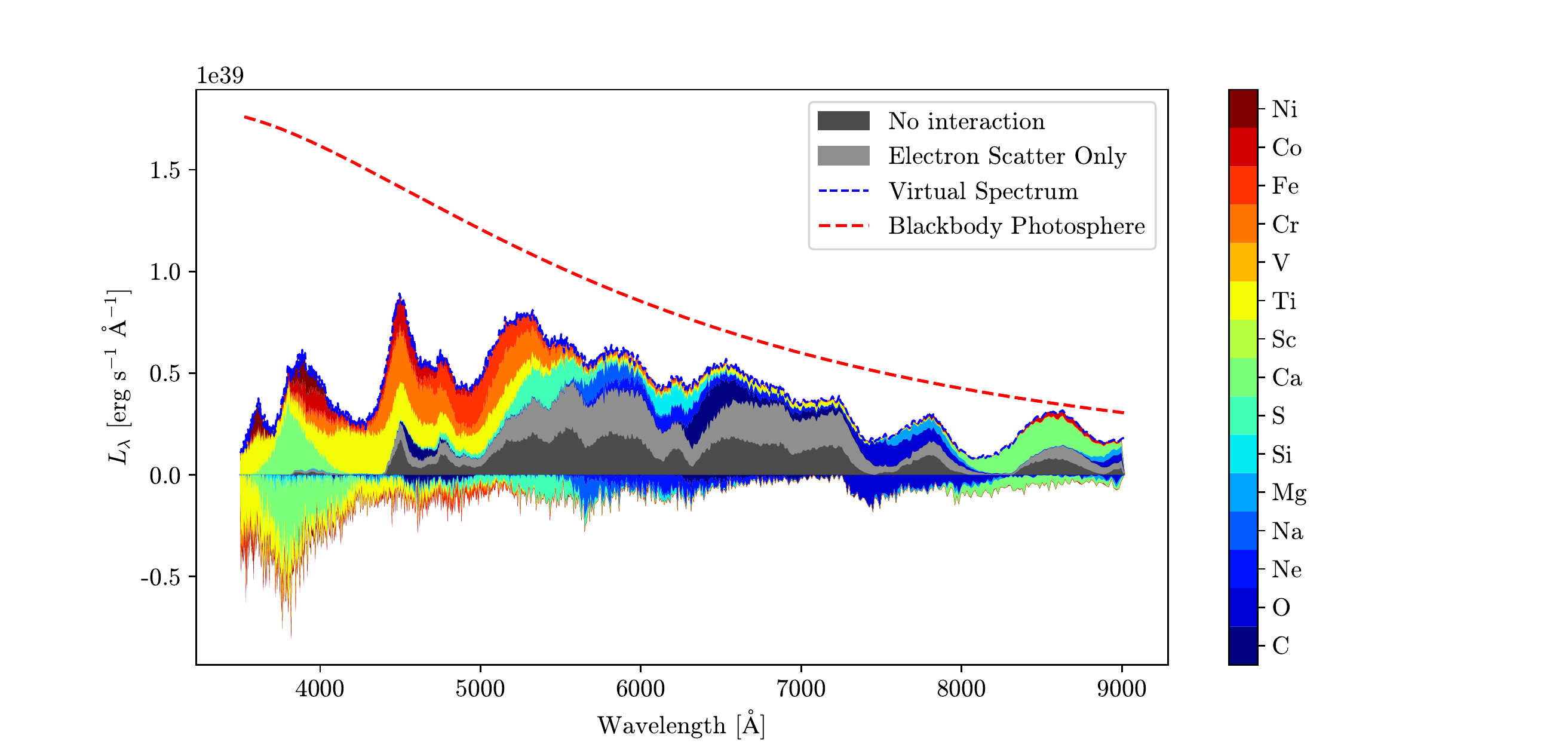}} 
    \subfigure[]{\includegraphics[width=0.45\textwidth]{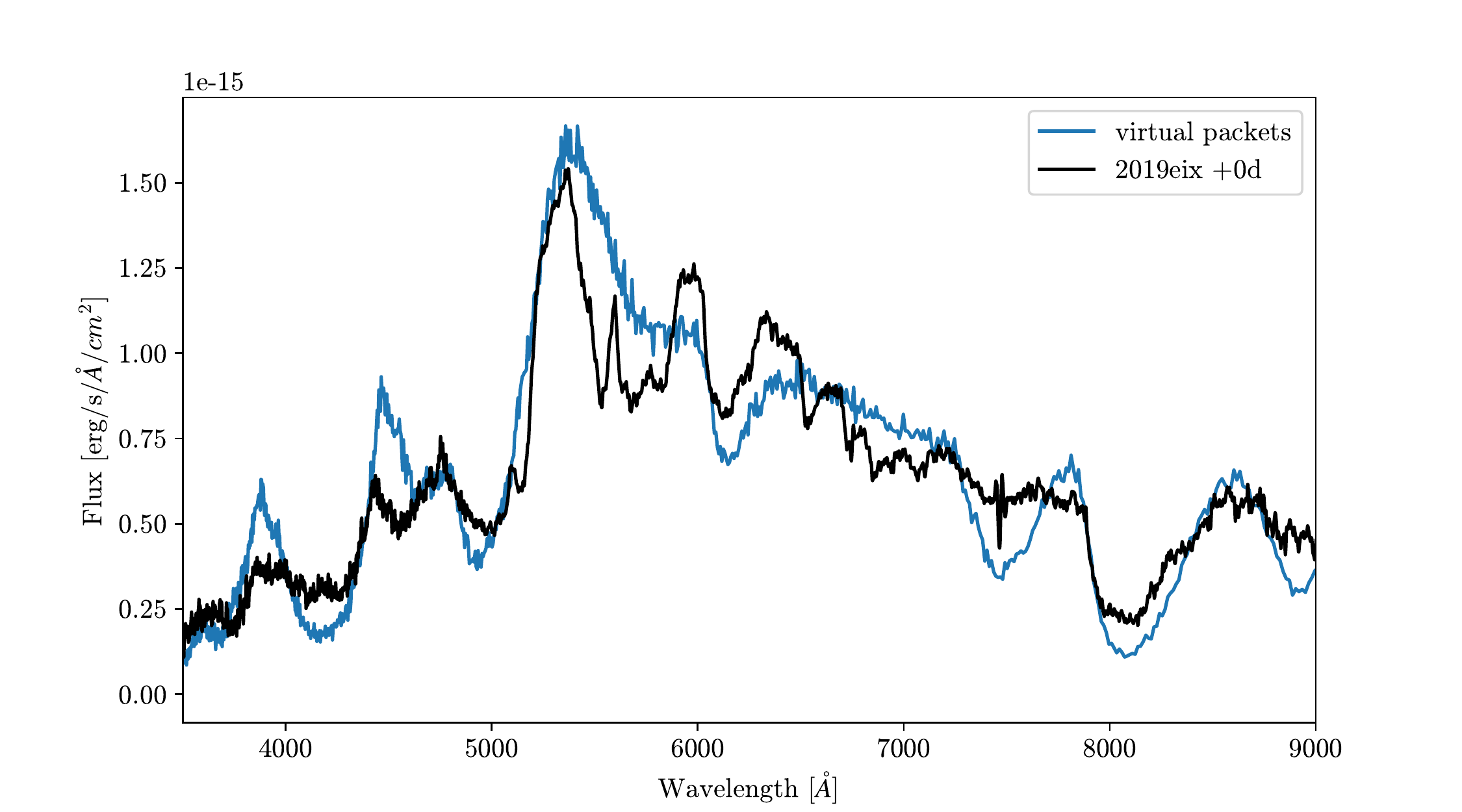}}
    \subfigure[]{\includegraphics[width=0.53\textwidth]{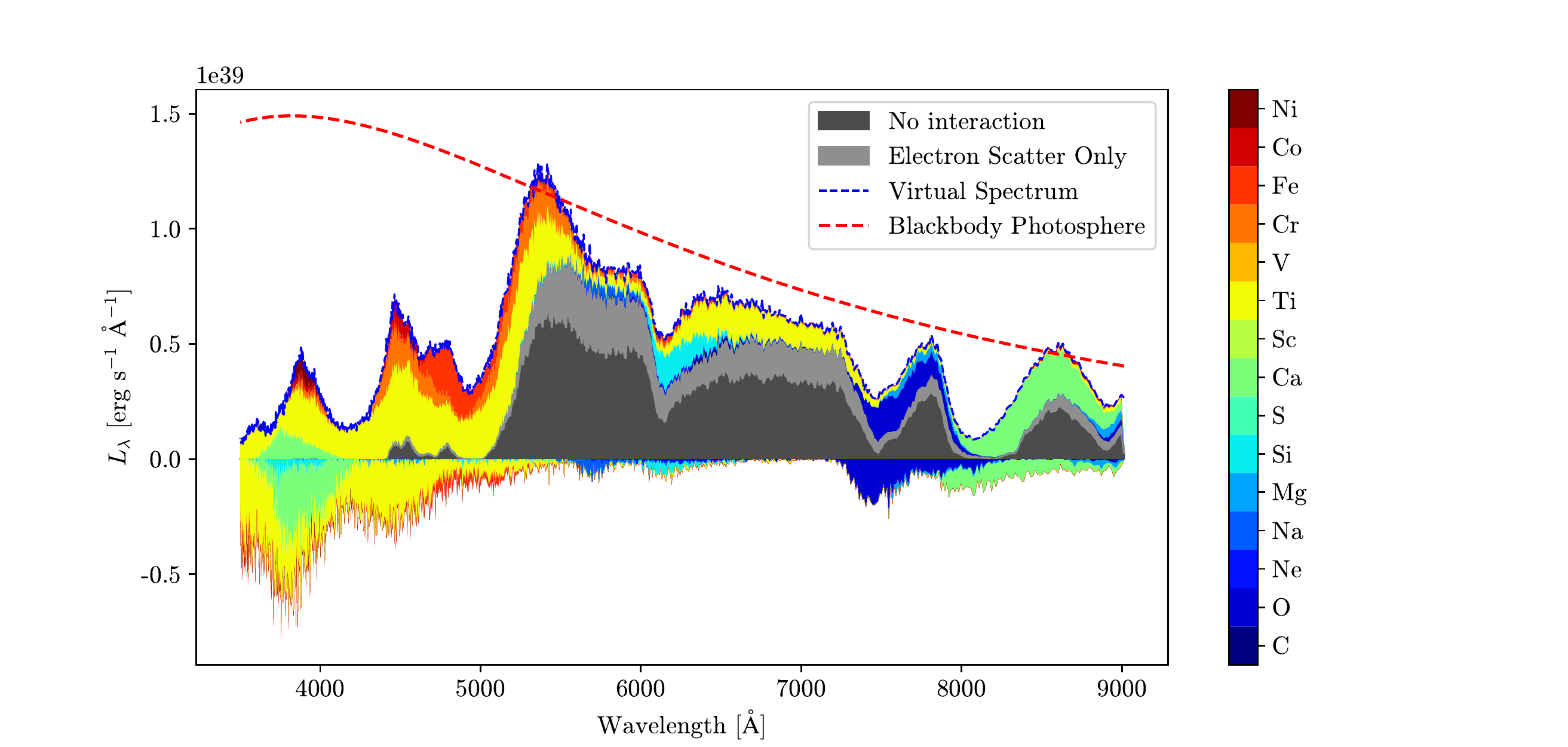}}
    \subfigure[]{\includegraphics[width=0.45\textwidth]{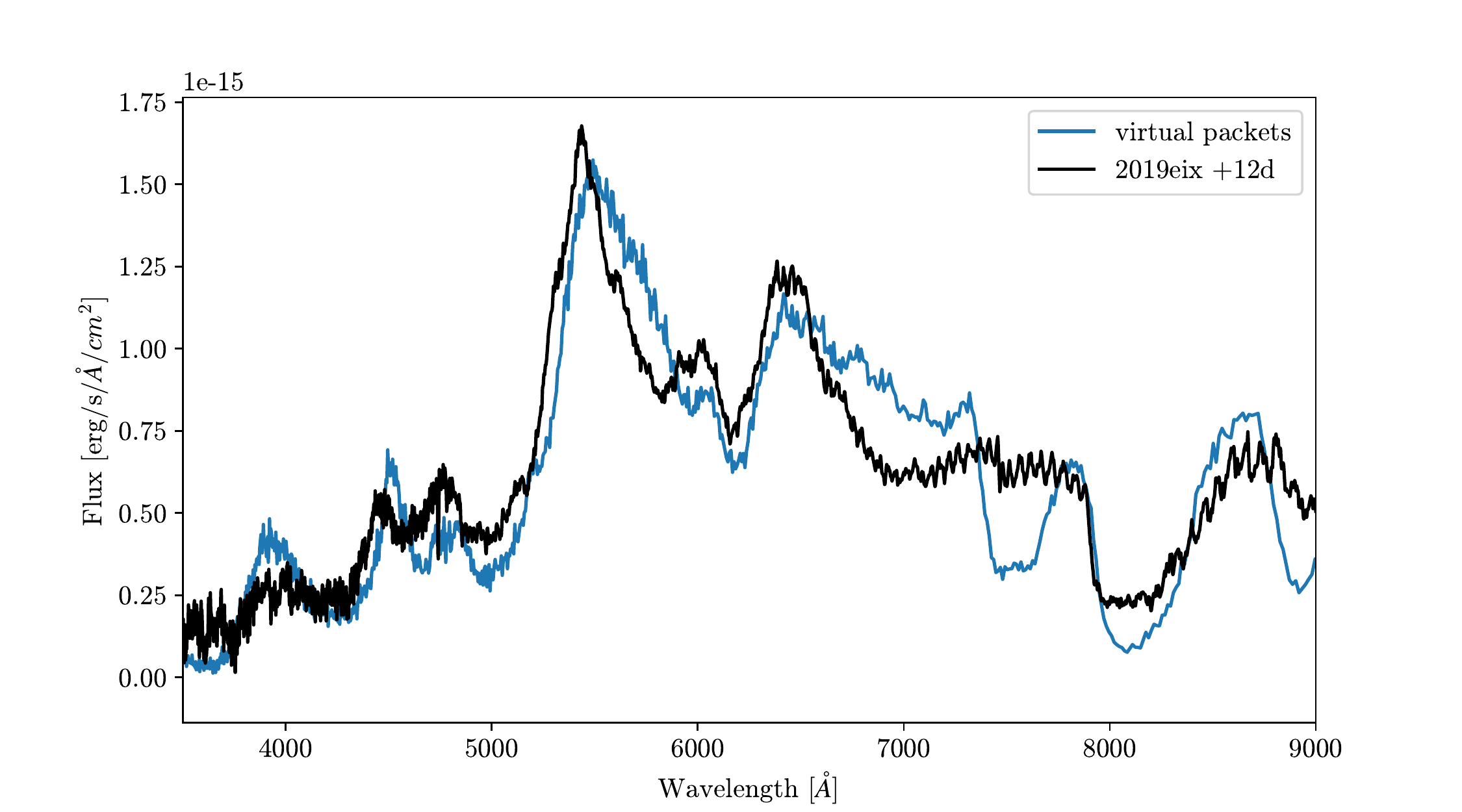}}
     \subfigure[]{\includegraphics[width=0.53\textwidth]{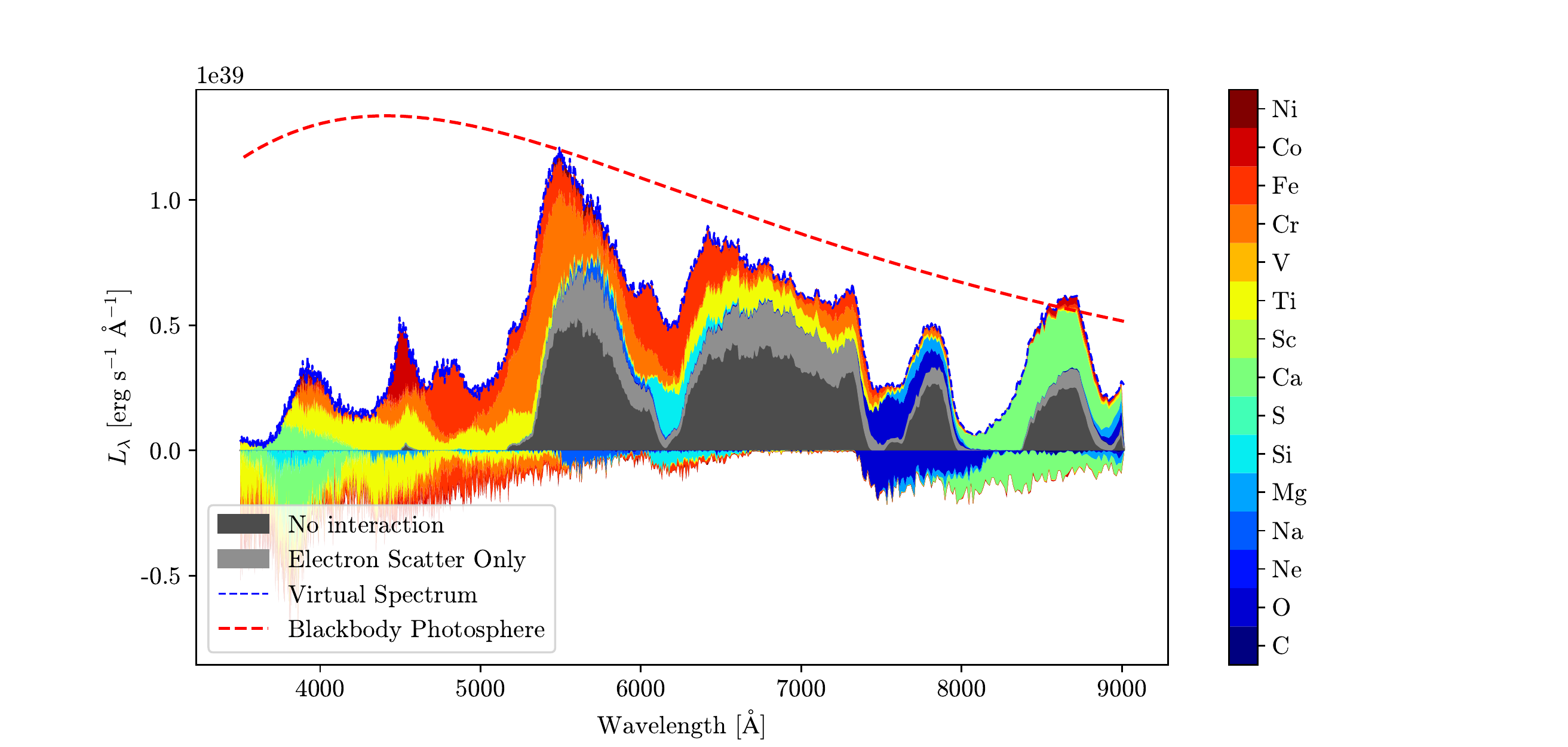}}
    \caption{We plot the best matched TARDIS models to SN 2019eix using SN 1994I SN Ic as a base model. On the left we have SN 2019eix and the final product of the TARDIS models while on the right it exhibits the elements responsible for its features. The negative values correspond to absorption contributions to the spectrum, while the positive values represent emission. TARDIS fails to hide the OI feature, but appears to reproduce most of the features overall.}
    \label{fig:tardis}
\end{figure*}

\subsection{Double-Detonation from He shell models} \label{sec:style}
SN 2019eix shares numerous similarities with double-detonation (DD) candidates from He shell Type Ia as shown in Figures~\ref{fig:lc_comp} and ~\ref{fig:spec_comp}. Spectroscopicaly, SN 2019eix shows significant red colors as well as suppression in emission lines below 5000 \AA; similar to SN 2016hnk \citep{Jacobson-galan2020}, SN 2018byg \citep{De19}, and SN 2016dsg starting from maximum following to 30 days after maximum light. The strong blanketing in the DD on the He shell scenario can be attributed to the large amount of iron group elements created in the outer ejecta from the He burning. The substantial reddening can be explained by the ashes produced by the He shell detonation causing redder colors throughout its evolution. Photometrically, SN 2019eix appears to be brighter and with a somewhat wider lightcurve than the other DD He shell candidates as shown in Figure \ref{fig:lc_comp}, with the exception of SN 2016hnk which shows a slower evolution than SN 2019eix post maximum. However, in terms of the color evolution SN 2016hnk seems to be the best match as shown in Figure \ref{fig:color_evol} from the extensive comparison with multiple SNe of various types.

Due to the similarities of SN 2019eix with this class we compare the spectra and the light curves to the double detonation models from \cite{Kromer10}, \cite{Sim12}, and \cite{Polin19}. \citet{Kromer10} explored observable properties of double detonation models. The simulations presented were carried out in 2D using radiative transport from \texttt{ARTIS}. To initialize models, they chose estimated values of temperature, central density of the CO core, and temperature and density at the base of the He layer. They ignited an initial He detonation in a single point at the base of the He shell and eventually created a shock wave that propagates into the core. These models have He shell masses ranging from 0.0035 to 0.0126 $\rm M_{\odot}$ previously considered by \cite{Fink10}. 

Similar to \cite{Kromer10}, \cite{Sim12} created their models using \texttt{ARTIS} and investigated small (0.45 $\rm M_{\odot}$ WD + 0.21 $\rm M_{\odot}$ He) and large (0.58 $\rm M_{\odot}$ + 0.21 $\rm M_{\odot}$ He) masses for both single and double detonation scenarios. The two methods investigated were the He detonation wrapping around the CO WD, modeled as the convergence of shocks in the first method, and the He detonation igniting an inward-propagating shock at the CO core's edge, called the Edge-lit core detonation in the second approach.
Additionally, \citet{Polin19} examined the explosions of WDs varying from 0.6 to 1.2 $\rm M_{\odot}$ with He shell masses of 0.01, 0.05, 0.08 $\rm M_{\odot}$. \cite{Polin19} found that thicker shell models showed early time flux excess, redder colors, and higher line blanketing in the UV through the blue regime of the spectrum. Their models were created using the Eulerian hydrodynamics code \texttt{Castro}. After the SN ejecta reaches homologous expansion, a multi-dimensional time dependent radiation transport code (\texttt{SEDONA}) is used to created the synthetic spectra and light curves.



\begin{figure}
\includegraphics[width=0.45\textwidth]{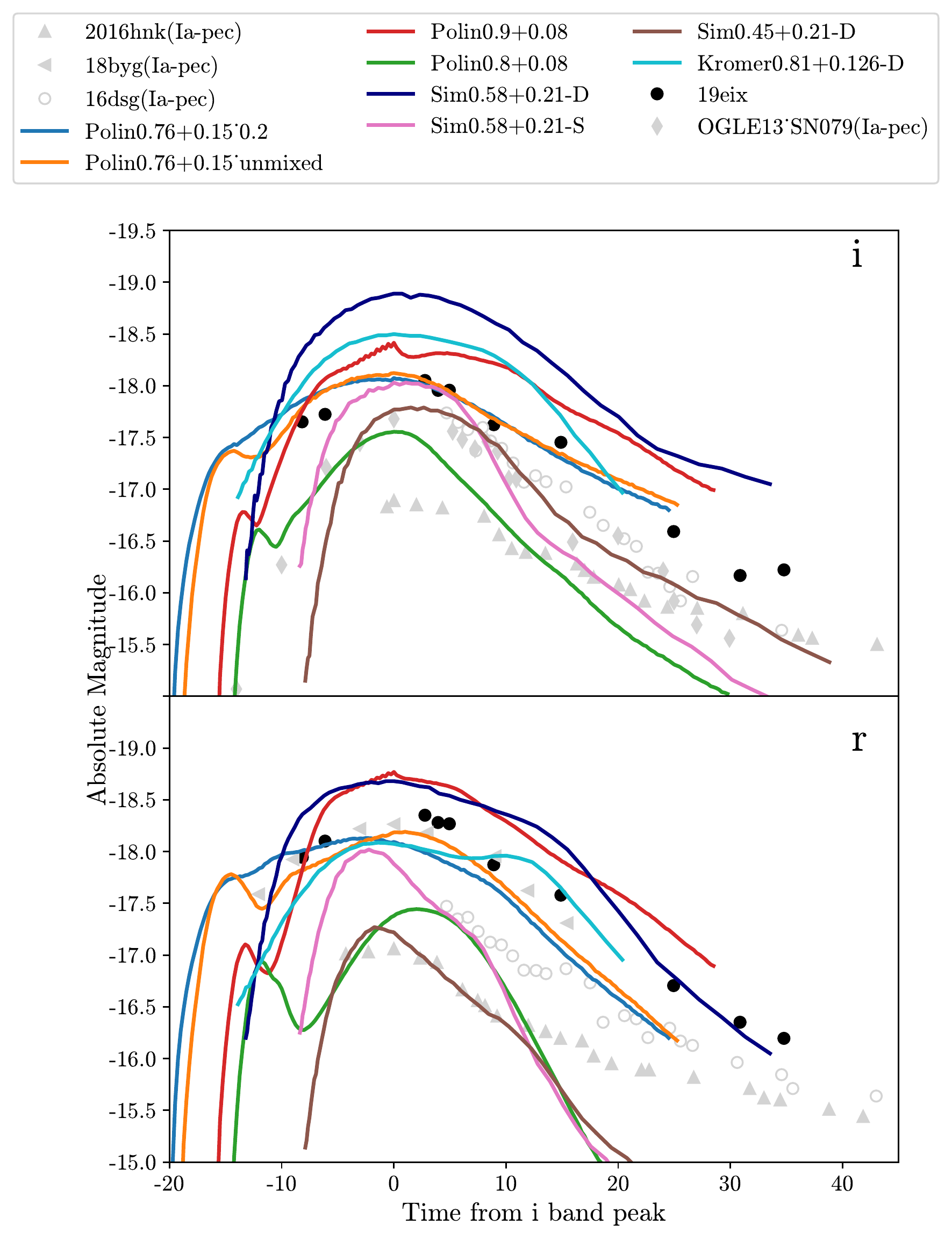}
\caption{Lightcurve comparison between SN 2019eix and other DD candidates to detonations of He-shell models. The phase is measured from the i/I-band maximum. Notice there is not a light curve that matches both the r and i the band, but Polin 0.76+0.15 unmixed is the closest match to SN 2019eix.
\label{fig:dd_lc}}
\end{figure}

In our study, the models we chose to compare to SN 2019eix are the double detonation models Polin0.9+0.08-D, Polin0.76+0.15-D, Polin0.76+0.15-0.2-D, Kromer0.81+0.126-D (model1), and Sim0.58+0.21-D (CSDD-S) as shown in Table \ref{tab:dd_models}. We chose these models as they appeared to be more consistent with our lightcurves and spectra of SN 2019eix. However, we also considered Polin0.8+0.08, Sim 0.45+0.21-D, and the only single detonation, Sim0.58+0.21-S. From Figure~\ref{fig:dd_lc}, we can see that photometrically, the more massive model with the thicker He shell such as Sim0.58+0.21-D (CSDD-S) overestimates the brightness in both bands. The other models Polin0.76+0.15-D, Polin0.76+0.15- 0.2-D, and Kromer0.81+0.126-D (model1) seem to be a better match to SN 2019eix, especially Polin0.76+0.15-D. Nonetheless, Polin0.76+0.15-D (unmixed) our best model underestimates the magnitude in the r band and declines much faster than SN 2019eix, but the i band is more consistent from maximum light up until 10 days after. Additionally, some of the challenges in modelling the lightcurves of SN 2019eix were attributed to the lack of early photometry. With earlier photometry, a more detailed comparison to double-detonation He-shell models that predict an early bump in the light curve as illustrated in Figure \ref{fig:dd_lc}.

\begin{figure*}
\begin{minipage}{\textwidth}
\centering
\includegraphics[width=0.7\textwidth]{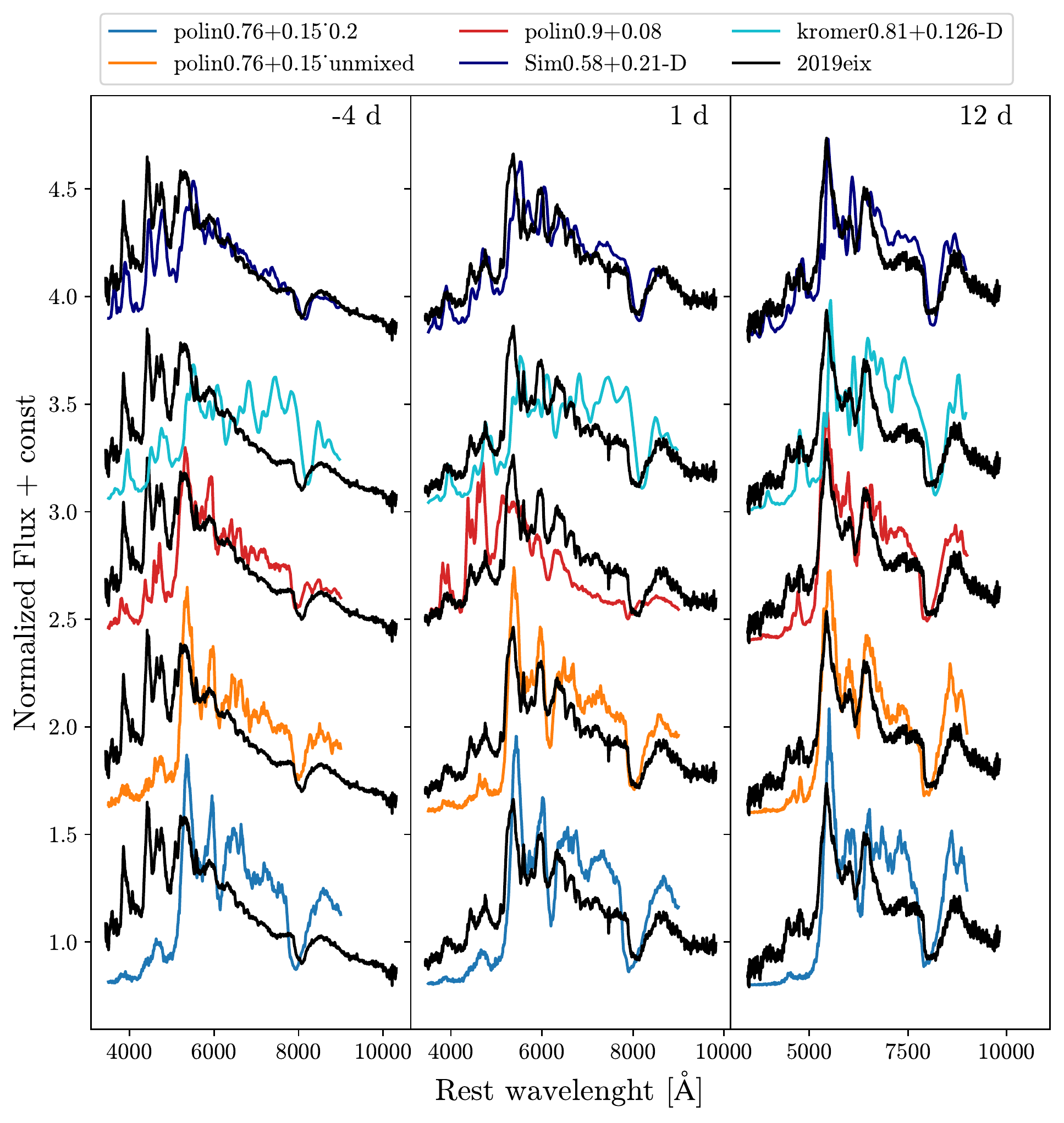}
\caption{ Spectroscopic comparison between SN 2019eix and double detonation models. The left panel shows the spectra about 4 days before maximum light, the middle panel is at 1 day after maximum light, and the right panel is 12 days after maximum light.
\label{fig:dd_spec}}
\end{minipage}
\end{figure*}

As mentioned, SN 2019eix shows substantial line blanketing in the blue side of the spectra. This is reproduced by the majority of the models at maximum and past maximum light as shown in Figure~\ref{fig:dd_spec}. Reproducing the data from the models consistently at various epochs was difficult. This is evident in Polin 0.9+0.08, which indicates that a reasonable result at one epoch was observed and not the others. Nevertheless, in Sim0.58+0.21-D (CSDD-S) we observe consistent behavior throughout the epochs, although the light curve of this model severely overestimates the flux. However, it is a known issue that \cite{Sim12} are brighter due to the pure He shell assumption, but in actuality the the He layer can be polluted by the WD core material. By contrast, the Polin0.76+0.15-D (unmixed) light curve model is the best match, but spectroscopically appears to be a mismatch before maximum light but overall a reasonable fit. 


Generally, it appears that the light curve models do not match SN 2019eix consistently. However, some of this mismatch could be attributed to the degree of mixing in the outer layer \citep{Kromer10, Shen10, Sim12, Polin19} in addition to the viewing angle \citep{Kromer10}. The models used for comparison are all from Local-Themodynamic Equlibrum which takes into effect about 20 days before maximum light. Therefore, it is important to compare the models to the data at early times. In order to fully compare the light curve at all epochs, we must use non-LTE models.

\begin{table*}
\begin{minipage}{\textwidth}
\begin{center}
\centering
\makebox[1 \textwidth][c]{       
\resizebox{0.8\textwidth}{!}{   
\hspace*{-9em}
\begin{tabular}{ccccc} 

\hline
\hline

Model & Original name & WD Mass & He Shell Mass & Reference \\
\hline
Polin0.76+0.15-0.2-D & &0.76 & 0.15 & a  \\
Polin0.76+0.15-D & &  0.76 & 0.15 & a \\
Polin0.9+0.08-D &  &  0.9 & 0.08 & a \\
Kromer0.81+0.126-D & Kromer model 1&  0.81 & 0.126 & b\\
Sim0.58+0.21-D &  Sim CSDD-S & 0.58 & 0.21 & c\\

\hline

\end{tabular}
} 
} 

\caption{\label{tab:dd_models} This table shows the models used for comparison with SN 2019eix. (a)\cite{Polin19}; (b) \cite{Kromer10}; (c)\cite{Sim12}. Note that Polin0.76+0.15-0.2-D and Polin0.76+0.15-D are similar, but differ since Polin0.76+0.15-0.2-D includes 0.2 $M_{\odot}$ mixing in the outer ejecta }
\label{table:dd_models}
\end{center}
\end{minipage}{}
\end{table*}

\section{Discussion} \label{sec:discussion}

SN 2019eix is an uncommon event. It has similar features to a SNe Ic including the lightcurve peak magnitude and decline rates, while having a strong suppression in the blue flux, red colors, a lack of an \ion{O}{1} feature, and a strong \ion{Ca}{2} feature unlike those usually seen in SNe Ic. Since it is unlike any known SN Ic, we also investigate other SN types including Calcium-rich transients and thermonuclear supernovae like subluminous supernovae and double detonation SNe Ia.

\subsection{Comparison to Type Ic and peculiar Ic} \label{sec:style}

In Figure \ref{fig:lc_comp}, we showed that the lightcurves for SN 2019eix are the most similar to SNe Ic in terms of peak magnitude (Table \ref{table:1}), bolometric luminosity (illustrated in Figure \ref{fig:bol_lum}), and decline rates (Table \ref{table:2}). However, from the spectral and color evolution standpoint we do not observe such consistency. Spectroscopically, before maximum SN 2019eix resembles SNe Ic with the exception of the absence of the \ion{O}{1} feature. However, we observe significant discrepancy in the spectra at later times as shown in Figure \ref{fig:mean_spec_comp} (albeit peculiar SNe Ic show some level of consistency as we will discuss below). Additionally, the color evolution of SN 2019eix appeared to be much more red than SNe Ic as depicted in Figure \ref{fig:color_evol}.

For peculiar SN Ic PTF12gzk, the color evolution at early times seems to agree with SN 2019eix (Figure\ref{fig:color_evol}), but a full comparison of colors was not conducted as as we lacked photometry at later times. Additionally, the spectrum of PTF12gzk before and after maximum matches SN 2019eix as it starts to develop a blue suppression after maximum light, but does not seem to be depleted of the \ion{O}{1} 7774 feature. PTF12gzk appears to also have more blended features than SN 2019eix suggesting higher velocities. \cite{Ben-Ami} reported PTF12gzk to exhbit large expansion velocities of $\approx$ 30000 km/s measured days after explosion from the Si II line. Additionally, the mass of ejecta was measured to be $\rm M_{ej}$ = 7.5 $\rm M_{\odot}$, about a factor of 3 larger than our estimates for SN 2019eix. This suggests a higher initial progenitor mass of 25-30 $\rm M_{\odot}$ \citep{Ben-Ami} and a lower-mass progenitor for SN 2019eix.


We considered the USSNe scenario due to the lack of \ion{O}{1} in SN 2019eix, as  these SNe can be stripped of their O in addition to H and He layers from a compact binary \citep{Woosley2020,Dessart2020}. We dismissed this scenario, since neither the spectra nor the lightcurves are comparable (eg. USSNe SN 2010X and SN 2005ek show no blue suppression and have \ion{O}{1} features unlike SN 2019eix, as shown in Figure \ref{fig:ult_st_comp}). The fast rise times of ultra-stripped supernovae \citep[typically between 5-10 days and magnitude of -16;][]{Takashi} and fast decline rates \citep[$\rm \Delta m_{15} \approx$ 3;][]{Drout13}, suggest a smaller progenitor than SN 2019eix.

 From observations we note that peculiar Type Ic's (PTF12gzk) could roughly produce spectra that look similar to SN 2019eix. Due to the peculiar spectral evolution of SN 2019eix we modeled the spectra using TARDIS. From our models we showed that spectra similar to SN 2019eix could be reproduced by using the abundance models from SN 1994I type Ic \citep{Hachinger12, Iwamoto94}, but again the \ion{O}{1} failed to be reproduced as discussed previously in Section \ref{sec:4.1}. Overall, the lightcurves appear to be consistent with SN Ic, however the spectral evolution appears to be atypical for this class, in particular the strong reddening, lack of \ion{O}{1}, and strong \ion{Ca}{2} at later times.

\subsection{Subluminous SNe Ia} \label{sec:style}

As shown in Figure \ref{fig:spec_comp}, SN 2019eix starts to resemble subluminous SNe Ia in its later spectra, although it is notably different before maximum. Photometrically, 91bg-likes reach peak magnitudes between -16.7 to -17.7 in the B band similar to SN 2019eix. Overall, despite the similarities between SN 2019eix and 91bg-likes in terms of the spectra and the absolute magnitude, they also display key differences. 

Some of these differences include: bluer colors as shown in Figure \ref{fig:color_evol} \citep[91bg-likes reach colors of $\rm (B-V)_{max} \approx 0.5-0.6$;][]{Taubenberger08,Sullivan} being much bluer than SN 2019eix ($\rm (B-V)_{max} \approx 1.8$); faster decline rates ($\rm \Delta m_{15}(B) \approx 1.8-2.1$) than SN 2019eix ($\rm \Delta m_{15}(B) \approx 1.2$); and spectroscopic disagreement in some absorption features (especially the \ion{Si}{2}, \ion{O}{1}, and \ion{Ca}{2} features and generally the spectra is not as suppressed in the blue as SN 2019eix); Thus the inferred $^{56}$Ni mass and ejecta mass for 91bg-like SNe were found to be between $\sim0.05$ and $\rm \sim 0.10 M_{\odot}$ \citep{Mazzali97,Sullivan} and $\rm \approx 0.5  M_{\odot}$ \citep{Stritzinger06}, respectively, significantly lower than SN 2019eix. For these reasons it is unlikely that SN 2019eix shares a similar progenitor as the 91bg-like objects. However, it is important to note that the $\rm M_{\rm ej}$ mass estimate could be overestimated for SN 2019eix, as it depends on the photospheric velocity and width of the light curve. 

\subsection{ Calcium-rich Transients } \label{sec:style}

\begin{figure}
\includegraphics[width=0.5\textwidth]{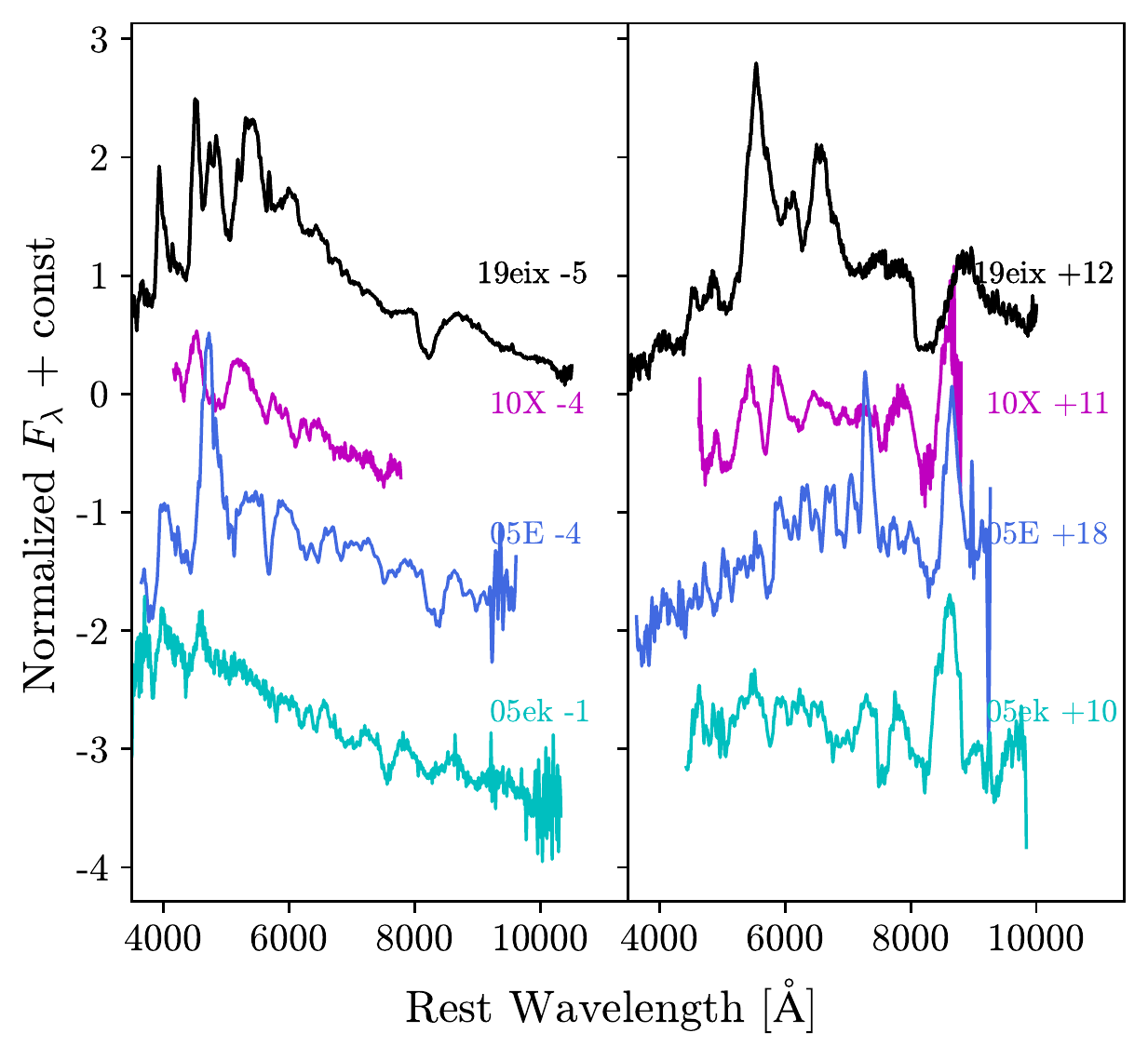}
\caption{A comparison between SN 2019eix and SN 2010X, SN 2005ek (ultra-stripped supernovae), and SN 2005E (cal-rich transient) the comparisons show the lack of compatibility between the two objects suggesting that SN 2019eix is not an ultra-stripped spectroscopically.
\label{fig:ult_st_comp}}
\end{figure}

Calcium-rich transients are peculiar as their explosion mechanisms is not well understood and a variety of scenarios can explain their observations. Due to strength of the \ion{Ca}{2} absorption feature on SN 2019eix, we briefly compare these transients to SN 2019eix, but rule them out due to significant incompatibilities. Calcium-rich SNe are mostly characterized by peak magnitudes of -14 to -16.5, with fast evolving light curves (much dimmer and faster evolving than SN 2019eix). From Figure \ref{fig:ult_st_comp} we see that SN 2005E (a Calcium-rich transient) is a mismatch to SN 2019eix in terms of its spectral evolution (e.g. it does not display the extreme suppression in the blue seen in SN 2019eix), but instead develops a [\ion{Ca}{2}] 7291, 7323 line 18 days after peak (not observed in SN 2019eix, but a common feature for these transients). Additionally, these transients also posses moderately red colors $\rm (B-V)_{max} \approx$ 0.6 mag \citep{Taubenberger}, unlike the significant red colors observed in SN 2019eix. The majority of these objects display low $^{56}$Ni and ejecta masses between 0.4-0.7 $\rm M_{\odot}$ \citep{Kasliwal12}. Due to the low luminosity, different color evolution, and incompatible spectra evolution; SN 2019eix is unlikely to share the same progenitor as these transients. 


\subsection{Double-detonation Type Ia }

The observations of He-shell double detonation candidates (OGLE-2013-SN-079, SN 2016hnk, SN 2016dsg, and 2018byg) display significant spectroscopic resemblance to SN 2019eix. Before maximum light, we do not observe much similarities and thus are not plotted. However, during and after maximum light the spectra of SN 2016dsg and SN 2018byg are found to be extremely suppressed in the blue and have a minimum to no absorption \ion{O}{1} feature, similar to SN 2019eix. During and after maximum light the spectra continue to have considerable line blanketing at $\lambda < 5000\rm{\AA}$ and show stronger \ion{Ca}{2} absorption as shown in Figure~\ref{fig:spec_comp}.

An explanation for the strong line blanketing in the blue for these double-detonation candidates is that the detonation of a helium shell on the surface of a C/O WD pollutes the outer layers of the ejecta. These IME and IGE burning products cause this blue blanketing and significant reddening \citep{Kromer10,Shen10,Sim12,Polin19}. The reddening is further illustrated in Figure~\ref{fig:color_evol}, where we see SN 2019eix sharing impressive similarities with SN 2016hnk as they both seem to be redder than the rest of the SN types. DD models also explain the strong \ion{Ca}{2} feature. This is a result of the He-burning creating a large amount of IME in the outer layers \citep{Fink10,Kromer10}. SN 2019eix develops \ion{Ca}{2} into a deep and high-velocity feature after maximum light, as shown in Figure \ref{fig:19eix_spec}. Therefore, a large amount of IME created in the DD models can explain such anomaly observed in SN 2019eix.


 We compare existing DD models to SN 2019eix in Figure \ref{fig:dd_lc} and Figure \ref{fig:dd_spec}. The models are roughly able to reproduce the lightcurve and spectral features. The best DD model of SN 2019eix is Polin 0.76+0.15 (unmixed), as it is able to be a better match to the lightcurves (although it still declines faster in the r band) and it's able to reproduce the most prominent features during and after maximum light (although it fails to match the features before maximum light). Notice in Figure \ref{fig:dd_spec} how the DD spectral models reproduce the lack of \ion{O}{1} feature, a pronounced feature in SN 2019eix. The physics of the absence of \ion{O}{1} feature in the case where the He shell is thick is justified by the ash of the He burning products created in the outermost layers of the ejecta. It is expected to cover the unburned oxygen from the core underneath the He products limiting them to a restricted velocity range \citep{Kromer10,Sim12,Polin19}. This could explain why the \ion{O}{1} feature for SN 2019eix is not observed until roughly 30 days after maximum.

Figure \ref{fig:lc_comp} shows that DD candidates show some level of agreement with SN 2019eix, particularly in the r-band. Existing candidates seem to be dimmer and faster declining than SN 2019eix (Figure \ref{fig:bol_lum}), although not many of these objects have been found in literature. Therefore, SN 2019eix remains a great candidate for the DD He shell model (with a possible WD mass close to 0.76 M$_{\odot}$ and a thick He-shell close to 0.15 M$_{\odot}$), as it is able to physically explain the abnormalities from the spectra and the color evolution unlike other classes. Therefore, for those reasons the DD scenario is the favored scenario for SN 2019eix.

\section{Conclusion} \label{sec:conclusion}

We summarize our primary observations of SN 2019eix below.

\begin{enumerate}
    \item SN 2019eix is intrinsically red compared to other SNe Ic. This is indicated by the lack of Na ID absorption in the photospheric spectrum and the color evolution of B-V 
    \item During and after maximum spectra show strong suppression in the blue
    \item Spectra show some of the strongest ever observed broad and deep \ion{Ca}{2} absorption and little to no \ion{O}{1} absorption
    \item Peak absolute magnitude reaches $M_{B} = -17.3$ with a decline parameter of $\Delta m_{15}(B) =1.2$
    \item Total ejecta and nickel masses are $2.5M_{\odot}$ and $0.09M_{\odot}$, respectively
\end{enumerate}

While the lightcurve decline rate and luminosity of SN 2019eix are consistent with SNe Ic, the spectra were challenging to reproduce. Notably, the lack of an \ion{O}{1} feature is inconsistent with both observed SNe Ic and our attempts to model it as such.


The thermonuclear double detonation scenario, on the other hand, is a better match to the data. The light curves from Polin 0.76+0.15 models seem to decline slightly faster than SN 2019eix and underestimate the luminosity for SN 2019eix in the r band, but have better agreement in the i band. However, spectroscopically we see SN 2019eix matches the spectral evolution of many DD candiates during and after maximum light as well as models. Moreover, the mystery of the O I is also explained by DD models along with the blueward supression. Conversely, most DD He-shell SNe candidates appear to come from elliptical galaxies and to be far away from the center of the galaxy, that is inconsistent with the host spiral galaxy of SN 2019eix. However, spiral galaxies also have an old stellar population and thus the DD scenario remains a possibility for SN 2019eix. Table \ref{table:ccvdd} summarizes the pros and cons between the core-collapse and the double-detonation thermonuclear events, favoring the latter. 

Because SN 2019eix is unable to be fully explained by either model (core-collapse or the DD He-shell) despite having a better physical explanation in the DD scenario (as it shares some of the peculiar spectral features and colors seen in SN 2019eix); there are some open questions in the DD scenario that point to future study. The light curve width is large, possibly implying a large ejecta mass. This could be inconsistent with a white dwarf progenitor. We thus classify this SN as a peculiar Type I SN. Due to the lack of early photometry (to constrain the rise time), nebular data (to observe whether it is Fe rich down to the core), and IR data (to check if there is unburnt He from the shell in the 10,500$\rm{\AA}$ line), we are unable to conclude with certainty that SN 2019eix is the result of a DD He-shell explosion. To get a better sense of the progenitor, further analysis of the SN 2019eix host galaxy star formation rate is essential to analyze whether the progenitor could have come from a massive star, unfortunately such analysis is beyond the scope of this paper. Nonetheless, SN 2019eix is an exciting target as it is arguably one of the rarest objects we have discovered so far, whether it is a double-detonation (being one of the handful of events ever discovered) or a core-collapse (being an extremely rare object with many unconventional features for this kind).

\vspace{5mm}
We are grateful for the National Science foundation (NSF) and the University of California, Santa Barbara (UCSB) for funding this project through the NSF-GSP-1911225 grant and the Central Campus Fellowship.

This research makes use of observations from the Las Cumbres Observatory network, in addition to the MARS ZTF alert broker developed by Las Cumbres Observatory software engineers.

This research has made use of the NASA/IPAC Extragalactic Database (NED) which is operated by the Jet Propulsion Laboratory, California Institute of Technology, under contract with NASA.

This research made use of \textsc{tardis}, a community-developed software package for spectral synthesis in supernovae \citep{Kerzendorf14, kerzendorf_wolfgang_2022_7428821}. The development of \textsc{tardis} received support from GitHub, the Google Summer of Code initiative, and from ESA's Summer of Code in Space program. \textsc{tardis} is a fiscally sponsored project of NumFOCUS. \textsc{tardis} makes extensive use of Astropy and Pyne.

The research by Y.D. is supported by NSF grants AST-2008108

M.W. is supported by the NASA Future Investigators in NASA Earth and Space Science and Technology grant (80NSSC21K1849).

M.M. is supported in part from NASA under the Swift GI program 1619152 (NASA grant No. 80NSSC21K0280), the Tess GI program G03267 (NASA grant No. 80NSSC21K0240), the ADAP program grant No. 80NSSC22K0486, and the HST GO program HST-GO-16178.007

\facilities{Las Cumbres Observatory (Sinistro), FTN (FLOYDS), Zwicky Transient Facility (ZTF)}

\software{\texttt{astropy} \citep{2013A&A...558A..33A,2018AJ....156..123A}, TARDIS \citep{Kerzendorf14}, SESNtemplate: \url{https://github.com/nyusngroup/SESNtemple}, Nugent's Spectral Templates: \url{https://c3.lbl.gov/nugent/nugent_templates.html}}

\appendix
\section{Appendix information}\label{sec:Appendix}
\begin{figure}[h]
\centering
\includegraphics[width=0.6\textwidth]{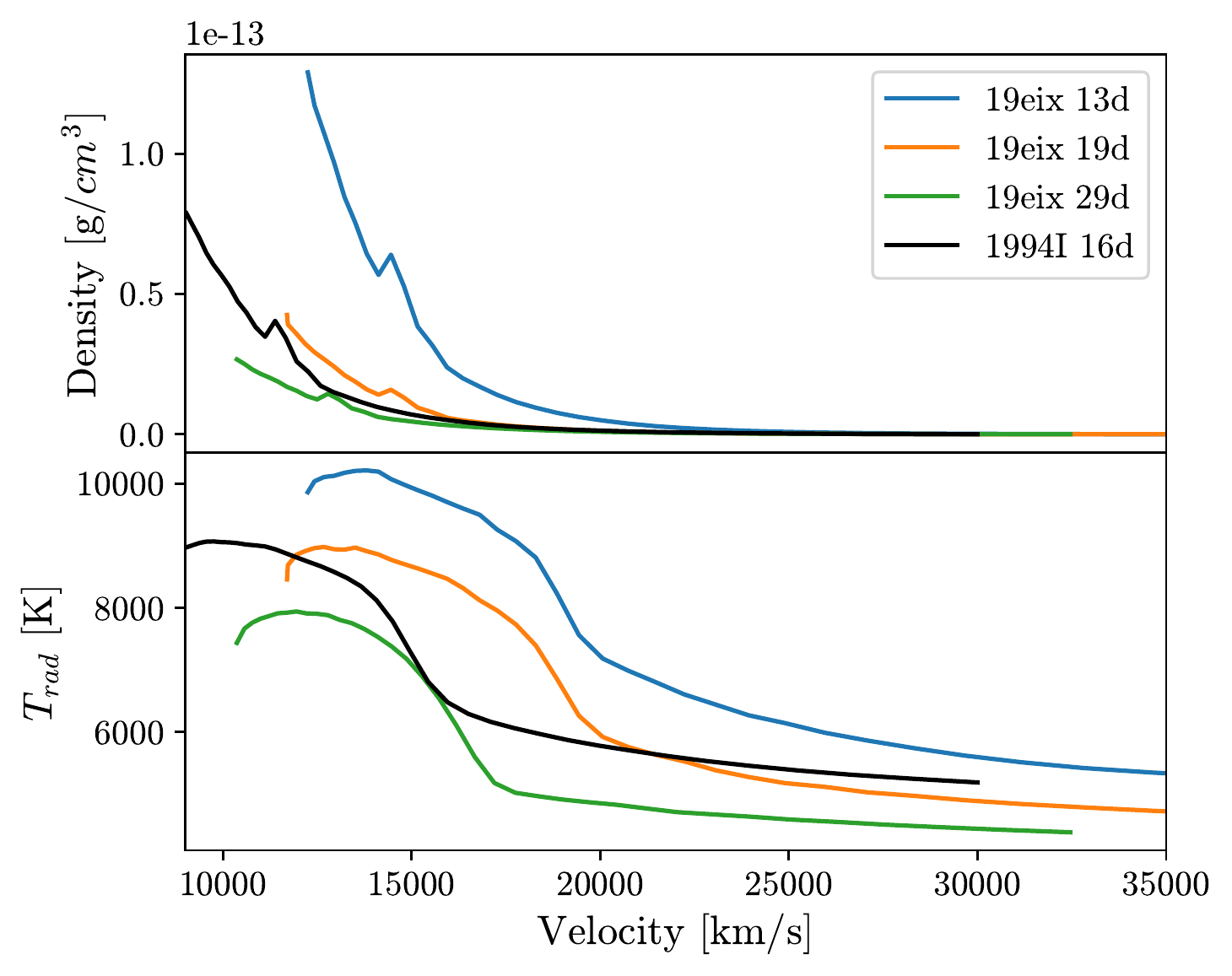}
\caption{The plot on the top shows the altered density profiles used for SN 2019eix for the different epochs we modeled from the base model SN 1994I. The bottom plot shows the adjusted temperature from the base model as well.
\label{fig:dens_prof_Tardis}}
\end{figure}

 The tables and plots are design to facilitate compare and contrast between the different SN classes/models to SN 2019eix. In Table \ref{table:summary_sne}, we provide additional information about the types of SNe and their various parameters for a more obvious comparison to SN 2019eix. For a better understanding of the spectra modeling conducted, in Figure \ref{fig:dens_prof_Tardis}, we show the the different input values we used for our TARDIS models such as density and temperature over the velocity space. In Table \ref{table:tardis_elem}, we provide the changes we have made to the abundances from the base model with respect to time from TARDIS. Finally, in Table \ref{table:ccvdd}, we created a table showing the main differences between core-collapse and the double-detonation for an easier comparison between the two.

\begin{table*}
\begin{minipage}{\textwidth}
\begin{center}
\centering
\makebox[1 \textwidth][c]{       
\resizebox{0.8\textwidth}{!}{   
\hspace*{-9em}
\begin{tabular}{cccccccccc} 

\hline
\hline

Model & time & velocity & X(Na) & X(Mg) & X(Si) & X(Ti)& X(Fe) & X(Ni)\\
\hline
1994I       & 16.0 &8900 & 4.0 & 10.0 & 0.7 & 0.08 & 0.03 & 1.8 \\
19eix Mod 1 & 13.0 & 12250 & 4.0 & 5.0 &3.0 & 0.4 & 0.16 & 1.8\\
19eix Mod 2& 19.0 & 11700 & 4.0 & 10.0& 0.7 & 2.0 & 0.5 & 2.2 \\
19eix Mod 3& 29.0 & 10370 & 0.7 &10.0 & 0.7& 0.08 & 0.9 & 2.6\\
\hline

\end{tabular}
} 
} 

\caption{\label{tab:table-name} This table shows the elements changed from the base model SN 1994I, the elements were slightly increased in Fe and Ni as time evolved. }
\label{table:tardis_elem}
\end{center}
\end{minipage}{}
\end{table*}

\begin{table*}
\begin{center}
\centering
\makebox[1 \textwidth][c]{       
\resizebox{0.4\textwidth}{!}{   

\begin{tabular}{ccc} 

\hline
\hline
Variables & Core-collpse & DD \\
\hline
Light curves & \checkmark & ? \\
$M_{\rm Ni}$ = 0.17 & \checkmark & ? \\
$M_{\rm ej}$ $>$ 2.5 & \checkmark & \xmark\\
no O I &  \xmark& \checkmark\\
Line blanketing  & \xmark & \checkmark \\
$\Delta m_{15}(B)$  & \checkmark  &  ?\\
Strong Ca II & \xmark &  \checkmark\\

\hline

\end{tabular}
} 
} 

\caption{\label{tab:table-name} This table summarizes the similarities and differences between SN 2019eix and the different explosion mechanisms. The entries with \textquote{?} represent measurements with not enough data or with large uncertainty. We do see that overall the double-detonation has stronger pros and weaker cons.   }
\label{table:ccvdd}
\end{center}
\end{table*}

\begin{table*}
\begin{center}
\begin{tabular}{ccccccc}
\hline
\hline
JD & epoch & B & V & g & i & r \\
\hline
2458608 &  -5 &             & 16.53(0.02) & 16.74(0.04) & 16.93(0.02) & 16.64(0.02) \\
2458610 &  -3 & 17.65(0.04) & 16.5(0.05)  & 17.03(0.02) & 16.86(0.1)  & 16.48(0.03) \\
2458619 &   6 & 18.37(0.19) & 16.53(0.07) & 17.38(0.02) & 16.53(0.03) & 16.23(0.02) \\
2458620 &   7 & 18.57(0.09) & 16.6(0.02)  & 17.55(0.01) & 16.62(0.03) & 16.3(0.01)  \\
2458621 &   8 & 18.43(0.16) & 16.61(0.04) & 17.6(0.04)  & 16.62(0.01) & 16.31(0.01) \\
2458625 &  12 & 18.86(0.17) & 17.1(0.05)  & 18.28(0.13) & 16.96(0.06) & 16.71(0.1)  \\
2458631 &  18 & 19.16(0.05) & 17.39(0.05) & 18.41(0.03) & 17.13(0.02) & 17.0(0.03)  \\
2458634 &  21 & 19.42(0.09) & 17.62(0.02) &             &             &             \\
2458641 &  28 & 19.63(0.02) & 18.08(0.01) & 18.99(0.03) & 17.99(0.03) & 17.87(0.01) \\
2458647 &  34 & 19.81(0.08) & 18.38(0.01) & 19.17(0.05) & 18.41(0.09) & 18.23(0.01) \\
2458651 &  38 &             & 18.67(0.06) & 19.22(0.06) & 18.36(0.04) & 18.38(0.03) \\
2458655 &  42 & 19.88(0.13) & 18.63(0.04) &             &             &             \\
2458659 &  46 &             &             & 19.41(0.04) &             &             \\
2458670 &  57 & 19.93(0.1)  & 19.1(0.08)  & 19.59(0.03) & 19.13(0.14) & 19.28(0.23) \\
2458673 &  60 &             & 19.48(0.12) &             &             & 19.43(0.06) \\
2458681 &  68 & 19.97(0.09) & 19.42(0.09) & 19.63(0.08) &             & 20.06(0.16) \\
2458689 &  76 & 20.34(0.09) & 19.55(0.19) & 19.93(0.07) & 19.89(0.21) & 20.15(0.03) \\
2458699 &  86 & 20.47(0.1)  & 20.05(0.03) & 20.34(0.05) & 20.41(0.09) & 20.58(0.1)  \\
2458707 &  94 &             & 20.1(0.17)  & 20.2(0.11)  & 20.68(0.11) & 20.95(0.15) \\
2458718 & 105 & 20.91(0.2)  & 20.81(0.12) & 20.64(0.15) & 21.05(0.38) &             \\
2458726 & 113 & 20.87(0.16) & 20.95(0.26) & 20.83(0.21) &             &             \\

\hline
\end{tabular}
\caption{Phase with respect to the V max where the dates are rounded up.}
\label{tab:data}
\end{center}
\end{table*}

\begin{table*}
\begin{center}
\begin{tabular}{ccccccc}
\hline
\hline
UT Date  & epoch & Telescope &  Instrument \\
\hline
2019-05-05  &-4 & Lick & KAST \\
2019-05-10  &+1 & FTN & FLOYDS \\
2019-05-14 &+5 & FTN & FLOYDS \\
2019-05-22  &+13 & FTN & FLOYDS \\
2019-06-01  &+23 & FTN & FLOYDS \\
2019-06-11 &+33 & FTN & FLOYDS \\
\hline
\end{tabular}
\caption{Phase with respect to the V max}
\label{tab:data}
\end{center}
\end{table*}


\vspace{10mm}

\clearpage
\bibliography{sample63}{}
\bibliographystyle{aasjournal}

\end{document}